\documentclass[twocolumn]{aastex63}

\usepackage{amsmath}
\usepackage{graphicx}
\usepackage{natbib}
\usepackage[caption=false]{subfig}

\newcommand{\mj}{$M_{\mathrm{J}}$} 
\newcommand{\rj}{$R_{\mathrm{J}}$}
\newcommand{\me}{$M_{\oplus}$}

\newcommand{\teff}{$T_{\rm eff}$}
\newcommand{\tint}{$T_{\rm int}$} 
\newcommand{\teq}{$T_{\rm eq}$}
\newcommand{\cp}{\citep} 
\newcommand{\ct}{\citet}

\newcommand{\tchem}{$t_{\rm chem}$}
\newcommand{\tmix}{$t_{\rm mix}$}
\newcommand{\kzz}{$K_{zz}$}
\newcommand{\gjf}{GJ 436b}
\newcommand{\gjt}{GJ 3470b}
\newcommand{\wasp}{WASP-107b}

\shorttitle{Transiting Planet Atmosphere/Interior Connection}
\shortauthors{Fortney et al.}

\begin{document}

\title{Beyond Equilibrium Temperature: How the Atmosphere/Interior Connection Affects the Onset of Methane, Ammonia, and Clouds in Warm Transiting Giant Planets}

\correspondingauthor{Jonathan J. Fortney}
\email{jfortney@ucsc.edu}

\author[0000-0002-9843-4354]{Jonathan J. Fortney}
\affiliation{Department of Astronomy \& Astrophysics, University of California,  Santa Cruz, CA 95064, USA}

\author[0000-0001-6627-6067]{Channon Visscher}
\affiliation{Chemistry \& Planetary Sciences, Dordt University, Sioux Center, IA 51250, USA}
\affiliation{Space Science Institute, Boulder, CO 80301, USA}

\author[0000-0002-5251-2943]{Mark S. Marley}
\affiliation{NASA Ames Research Center Moffett Field, Mountain View, CA 94035, USA}

\author[0000-0003-1150-7889]{Callie E. Hood}
\affiliation{Department of Astronomy \& Astrophysics, University of California,  Santa Cruz, CA 95064, USA}

\author[0000-0002-2338-476X]{Michael R. Line}
\affiliation{School of Earth \& Space Exploration, Arizona State University, Tempe AZ 85287, USA}

\author[0000-0002-5113-8558]{Daniel P. Thorngren}
\affiliation{Institute for Research on Exoplanets, Université de Montréal, Montréal, Québec, H3T 1J4, Canada}

\author[0000-0001-9333-4306]{Richard S. Freedman}
\affiliation{NASA Ames Research Center Moffett Field, Mountain View, CA 94035, USA}
\affiliation{SETI Institute, Mountain View, CA 94043, USA}

\author{Roxana Lupu}
\affiliation{BAER Institute, NASA Research Park, Moffett Field, CA 94035, USA}

\begin{abstract} The atmospheric pressure-temperature profiles for transiting giant planets cross a range of chemical transitions.  Here we show that the particular shape of these irradiated profiles for warm giant planets below $\sim$1300\,K lead to striking differences in the behavior of non-equilibrium chemistry compared to brown dwarfs of similar temperatures.  Our particular focus is H$_2$O, CO, CH$_4$, CO$_2$, and NH$_3$ in Jupiter- and Neptune-class planets.  We show the cooling history of a planet, which depends most significantly on planetary mass and age, can have a dominant effect on abundances in the visible atmosphere, often swamping trends one might expect based on \teq\ alone.  The onset of detectable CH$_4$ in spectra can be delayed to lower \teq\ for some planets compared to equilibrium, or pushed to higher \teq. The detectability of NH$_3$ is typically enhanced compared to equilibrium expectations, which is opposite to the brown dwarf case.  We find that both CH$_4$ and NH$_3$ can become detectable at around the same \teq\ (at \teq\ values that vary with mass and metallicity) whereas these ``onset'' temperatures are widely spaced for brown dwarfs. We suggest observational strategies to search for atmospheric trends and stress that non-equilibrium chemistry and clouds can serve as probes of atmospheric physics.  As examples of atmospheric complexity, we assess three Neptune-class planets GJ 436b, GJ 3470b, and WASP-107, all around \teq$=700$ K. Tidal heating due to eccentricity damping in all three planets heats the deep atmosphere by thousands of degrees, and may explain the absence of CH$_4$ in these cool atmospheres.  Atmospheric abundances must be interpreted in the context of physical characteristics of the planet.  

\end{abstract}

%\keywords{planets, atmospheres}

\section{Introduction}\label{intro}
\subsection{Atmospheric Characterization}
Even 25 years after the discovery of gas giant exoplanets \citep{Mayor95} we are still in our infancy in characterizing the atmospheres of these worlds.  Over the past two decades, astronomers have made fantastic strides to obtain spectra of exoplanets, but we still have much to do.  In the realm of transiting planets, observers have often been hindered by instruments aboard space- and ground-based telescopes that were never designed for precision time series spectrophotometry.  Even as dozens of planets have been seen in transmission spectroscopy \citep[e.g.,][]{Sing16} and occultation spectroscopy or photometry \citep[e.g.,][]{Kreidberg14b,Garhart20} our ability to understand the physics and chemistry of hydrogen-dominated atmospheres has been limited, principally by low signal-to-noise observations and limited wavelength coverage.  On the side of the directly imaged planets, telescopes like Keck, VLT, and Gemini have allowed more robust atmospheric spectroscopy, but with a sample size that is so far limited in number \citep[e.g.,][]{Konopacky13,Macintosh15,Lacour19}. 

It is with brown dwarfs, now numbering over 1000, with temperatures down to 250 K \citep{Luhman14, Skemer16} where robust atmospheric characterization has taken place over the past 25 years.  The major transitions in atmospheric chemistry and cloud opacity have now been unveiled \citep{Burrows01,Kirkpatrick05,Helling14,Marley15}, although major open questions still exist on the role of clouds in shaping the spectra across a range of \teff\ and surface gravity.  However, it should be clear that relying solely on the classic ``stellar'' fundamental quantities of \teff, log $g$, and metallicity has already shown its faults for these objects.  For instance, time-variability can reach tens of percent, and effects due to rotation rate \citep{Artigau18} and viewing angle have now been seen as important to take into account for atmospheric characterization \citep{Vos17}.

To understand the atmospheres of giant planets we will certainly need a \emph{larger} sample size than the brown dwarfs, for a similar level of understanding, as planets have many additional complicating factors \citep{Marley07b}.  For instance, substantial recent work has gone into assessing the \emph{Spitzer} IRAC 3.6/4.5 colors of cooler transiting planets, in order to better assess atmospheric metallicity and the role of CH$_4$ and CO absorption \citep{Triaud15,Kammer15,Wallack19,Dransfield20}.  The wide diversity of colors at a given \teq, much wider than is seen in brown dwarfs at a given \teff\ \citep{Beatty14,Dransfield20}, has been interpreted as needing a large dispersion in atmospheric metallicity and potentially C/O ratio.

Planets present additional complicating physics, such as heating from above, across a range of incident stellar spectral types \citep{Molliere15}, in addition to a range of UV fluxes.  The planets will have diverse day-night contrasts and circulation regimes, likely with very wide range of atmospheric metallicities \citep{Fortney13,Kreidberg14b} and non-solar abundance ratios \citep{Oberg11,Madhu14,Espinoza17}.  The cooling of the interiors of giant planets -- even the cooler giant planets not affected by the hot Jupiter radius anomaly -- is also still not fully understood \citep[e.g.,][]{Vazan15,Berardo17}

Key science goals of the \emph{James Webb Space Telescope} (\emph{JWST}) and \emph{ARIEL} are to obtain spectra of a wide range of planetary atmospheres \citep{Beichman14,Greene16,Tinetti18}.  In the realm of transiting giant planets, which have predominantly accreted their atmospheres from the proto-stellar nebula, one aspect of this science will be characterizing planets over a wide range of temperatures, to sample a wide range of transitions in atmospheric chemistry and cloud formation.  A significant amount of previous theoretical and modeling work have gone into trying to predict and understand trends in the atmospheres of these planets, going back to important early works such as \citet{Marley99} and \citet{Sudar00}, supplemented by later works like \citet{Fortney08a}, \citet{Madhu11b}, and \citet{Molliere15}.  Most of these papers have pointed to planetary equilibrium temperature, \teq, as the dominant physical parameter that determines atmospheric physics and chemistry, somewhat akin to \teff\ in stars.  While there are good reasons to think that this is indeed true, there are equally good reasons to think that \teq\ is only a starting point, and that other physical parameters can have a crucial effect on determining the atmospheric spectra that we will see.

Of course \teq\ is only part of the energy budget, and it is well-understood that $T_{\mathrm{eff}}^4 = T_{\mathrm{eq}}^4 + T_{\mathrm{int}}^4$, with \tint\ parameterizing the intrinsic flux from the planetary interior, and \teq\ from thermal balance with the parent star.  In Jupiter, for instance, \teq\ and \tint\ are similar, with neither dominating the energy budget \citep{Pearl91,Li18}.  Recently, \citet{Thorngren19,Thorngren20} pointed out that the radii of ``hot'' and ``warm'' Jupiter population can be used to assess the intrinsic flux coming from planetary interiors.  Often Jupiter-like values of \tint\ (100 K) had been chosen for convenience, but the inflated radius of a typical hot Jupiter goes hand-in-hand with a hotter interior and much higher \tint\ values (assuming convective interiors).

This work gives us the ability to better assess the depth of the radiative-convective boundary (RCB) in these strongly irradiated planets.  A key finding of \citet{Thorngren19} was the \tint\ values are typically larger (sometimes much larger) than previous expectations, which moves the RCB to lower pressures.  A higher \tint\ can remove or weaken cold traps in these atmospheres, which can alter atmospheric abundances and the depth at which clouds form.  Much additional work needs to be considered for these hot planets, perhaps much of it in the 3D context, given the large day-night temperature contrasts \citep{Parmentier18}.

The role of the current paper is to serve as a complement, of sorts, and extension to, the work of \citet{Thorngren19}, but mostly for cooler planets.  For planets below \teq$\sim$~1000 K, a wide range of chemical and cloud transitions should occur \citep{Marley99,Sudar00,Morley12}.  What is not as appreciated, however, is that temperatures in the deeper atmosphere, which are typically not visible, can play as large a role, or even a larger role, in determining atmospheric abundances as the visible atmosphere, which is dominated by absorbed starlight.

The temperatures of the deep atmosphere, while typically not measureable, can be constrained in a variety of ways.  Observationally, flux from the deep interior can potentially be seen at wavelengths where the opacity is low (``windows'').  This has been constrained for GJ 436b emission photometry \citep{Morley17a}, and could potentially be done for a small number of other planets \citep{Fortney17}.  Another is cold-trapping gases into condensates via crossing a condensation curve in the deep atmosphere \citep{Burrows07c,Fortney08a,Beatty19,Thorngren19,Sing19}. 

As was done in \citet{Thorngren19}, the planetary radius can be used as a constraint, with assumptions about interior energy transport.  Planetary thermal evolution/contraction models aim to understand the cooling of the planetary interior with time \citep[e.g.,][]{Fortney07a,Baraffe08}.  Furthermore, there are planets for which thermal evolution models can be made more uncertain -- those that are undergoing tidal eccentricity damping.  If this energy is dissipated in the planet's interior, the temperature of the deep atmosphere can be significantly enhanced compared to simple predictions.  Lastly, one can assess the role of disequilibrium chemistry tracers.  Recently, \citet{Miles20} have used observations of disequilibrium CO in cold brown dwarfs to understand atmospheric dynamics and temperature structures.  They constrain the rate of atmospheric vertical mixing as a function of \teff, providing strong evidence for a detached radiative zone, below the visible atmospheres, long predicted in these atmospheres \citep{Marley96, Burrows97}. Is is these disequilibrium tracers which we turn to next, in more detail.

\begin{deluxetable*}{ccccccc}
\tabletypesize{\footnotesize}
\tablenum{1}
\tablecaption{Guide to Model Parameters\label{tab:guide}}
\tablewidth{0pt}
\tablehead{
\colhead{Fig.} & \colhead{\teq (K)} & \colhead{\tint (K)} & \colhead{$M_{\textrm{J}}$} & \colhead{$g$ (m s$^{-2}$)} & \colhead{$m$} & \colhead{age (Gyr)}
}
%\decimalcolnumbers
\startdata
1 & 710 & 60, 100, 200, 300, 400 & 1 & 25 & 10$\times$ \\
4, 23 & 710 & 52, 77, 117, 182, 333 & 0.1, 0.3, 1, 3, 10 & 5.8, 9.8, 24, 65, 225 & 10$\times$ & 3\\
7, 13 & 1120 to 180 &  75 & 0.3 & 10 & 10$\times$ & 3\\
9, 15 & 870, 380, 180 & 52, 117, 333 & 0.1, 1, 10 & 5.8, 24, 225 & 10$\times$ & 3 \\
11, 17 & 710 & 501, 383, 283, 212, 156, 117, 84 & 1 & 13, 16, 19, 21, 23, 24, 26  & 3$\times$ & {0.01, 0.03, 0.1, 0.3, 1.0, 3.0, 10.0}\\
19 & 870, 380 & 52, 117, 333 & 0.1, 1, 10 & 5.8, 24, 225 & 1, 3, 50$\times$ & 3 \\
\enddata
\tablecomments{In each figure, a range of planetary models is considered explored across different planetary parameters.  The metallicity factor $m$ is defined as $m=10^{[\textrm{Fe/H}]}$.}
\end{deluxetable*}
\normalsize

\subsection{``Hidden" Atmospheric Chemistry}
Due to non-equilibrium chemistry via vertical mixing, deep atmosphere temperatures can matter as much as temperatures in the visible atmosphere in determining observable abundances.  This well-understood process affects abundances when the mixing timescale for a parcel of gas, $t_{\rm mix}$, it shorter than the chemical conversion timescale, $t_{\rm chem}$, for a given chemical reaction.  Well-studied reactions are CO to CH$_4$ and N$_2$ to NH$_3$.  These timescales can be so long that the gas in the visible atmosphere (at say, 1 mbar) will be representative of pressure-temperature (\emph{P--T}) conditions at $\sim$1-1000 bar, as we will readily show.  The effects of non-equilibrium chemistry on the atmospheric abundances and resulting spectra in giant planet (both solar system and extrasolar) and brown dwarf atmospheres have previously been extensively studied \citep{Fegley96,Saumon03,Saumon06,Visscher10,Visscher11,Moses11,Madhu11b,Venot12,Moses13,Miguel14,Zahnle14,Mola19,Venot20,Miles20,Mola20} and here we will not break new ground on the chemistry.  Rather, following the carbon and nitrogen chemistry work of \citet{Zahnle14}, we will point out several novel complexities that arise when applying non-equilibrium chemistry to the quite inhomogeneous exoplanet population. 
Given the very large uncertainties in vertical mixing speeds, in particular for these irradiated atmospheres that are mostly radiative rather than convective (where mixing length theory could plausibly be used), in addition to uncertainties in thermal evolution models, as well as the currently unknown atmospheric metal-enrichments, we will show that a very wide range of behavior should be expected.  For instance, one should not expect a single transition temperature in \teq\ from CO--dominated to CH$_4$--dominated atmospheres, an area of active study already with \emph{Hubble} and \emph{Spitzer} \citep{Stevenson10,Morley17a,Kreidberg18,Benneke19}.

\begin{figure}[hbp!]
\includegraphics[clip,width=1.0\columnwidth]{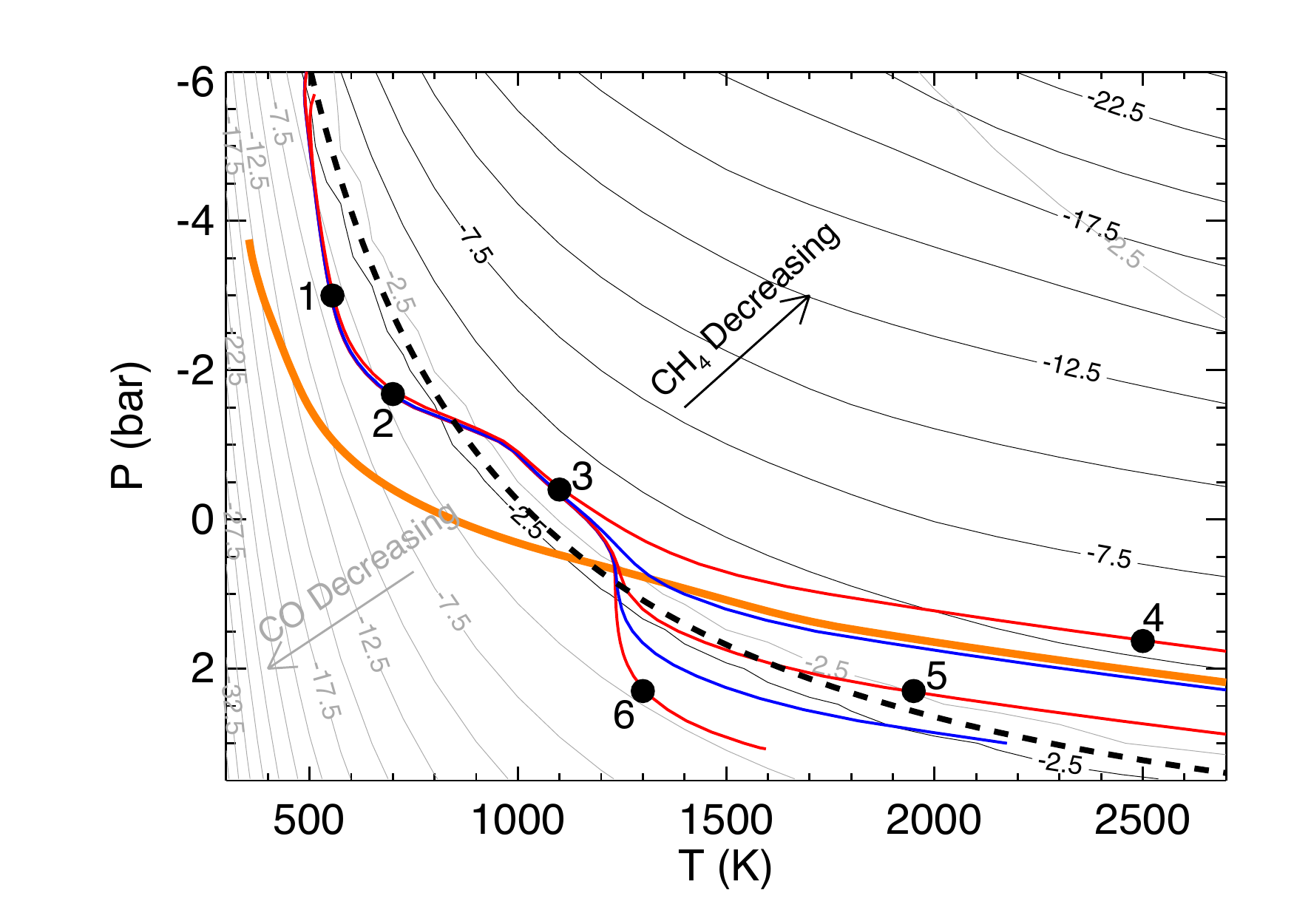}
\caption{Model pressure-temperature profiles for a $10\times$ solar atmosphere at 0.15 AU from the Sun.  The five profiles all have \teq = 710 K and show (alternating red and blue) five values of \tint, at 60, 100, 200, 300, and 400 K and a Jupiter-like gravity of 25 m s$^{-2}$. Also shown in thick orange is a \teff\ of 1000 K brown dwarf with a gravity of 1000 m s$^{-2}$.  Equal-abundance contours for CH$_4$ are shown in dark gray, and show the log (base 10) of volume mixing ratios of CH$_4$ that fall off by many orders of magnitude towards the upper right. Correspondingly, light gray contours show the same for CO, toward the lower left, where CH$_4$ is the dominant absorber.  CO and CH$_4$ have an equal abundance at the dashed thick black curve.  These mixing ratio contours assume equilibrium chemistry.  The numbered black dots are called out specifically in the text.
\label{profile}}
\end{figure}

We can first look at an illustrative example of why vertical mixing from different atmospheric depths can strongly affect observed abundances and spectra, by exploring the behavior of CO, CH$_4$, and H$_2$O.  Figure \ref{profile} shows the atmospheric pressure-temperature (\emph{P--T}) profile for a planet at 0.15 AU from the Sun, with \teq$=710$ K.  Five models are shown, with decreasing \tint, leading to cooler interior adiabats.  Underplotted in light gray are curves of constant volume mixing ratio (mole fraction) for CO, to the lower left, following the chemical equilibrium calculations of \citet{Visscher10} and \citet{Visscher12}.  Underplotted in dark gray is the same for CH$_4$, to the upper right.  The dashed thick black curve shows the equal-abundance boundary, where the mixing ratio of CO$=$CH$_4$:
\begin{equation}
\log_{10} P \approx 5.05 - 5807.5/T + 0.5\textrm{[Fe/H]},
\end{equation}
for $P$ in bar, $T$ in K, and [Fe/H] as the metallicity \cp{Visscher12}.  When we turn to nitrogen chemistry in Section \ref{nitrogen}, we will use the analogous N${_2}$=NH$_3$ equal abundance curve:
\begin{equation}
\log_{10} P \approx 3.97 - 2721.2/T + 0.5\textrm{[Fe/H]}
\end{equation}

Numbered black dots in Figure \ref{profile} have been placed along the profiles.  Point $1$ is at 1 mbar, a pressure that would be readily probed in transmission spectroscopy.  Point $2$ is at 700 K, where the local temperature is equal to \teff, a good representation of the mean thermal photosphere in emission.  Points $1$ and $2$ are in the CH$_4$-dominated region, with point $2$ having $\sim$~10$\times$ more CO.  Moving down to point $3$, all profiles are now in the CO-dominated regime, where the CH$_4$ abundance falls off dramatically with temperature.  Point $4$ is deeper in the atmosphere along the hottest adiabat, in the CO-rich region, with a decrease in CH$_4$ compared to point $3$.  Points $5$ and $6$ are along cooler adiabats, with $5$ having abundances quite similar to point $3$.  Point $6$ is quite interesting, in that, while it is in the deep part of the atmosphere, it is clearly within the CH$_4$-dominant region, and has the same CH$_4$ and CO abundances as point $2$.  This complexity should be contrasted with the profile of a \teff $=1000$ K, log $g$=5 brown dwarf, plotted in thick orange.  For the brown dwarf, as a parcel of gas moves along from high pressure to low, there is a monotonic increase in CH$_4$ and decrease in CO.

\begin{figure}[hbp!]
\includegraphics[clip,width=1.0\columnwidth]{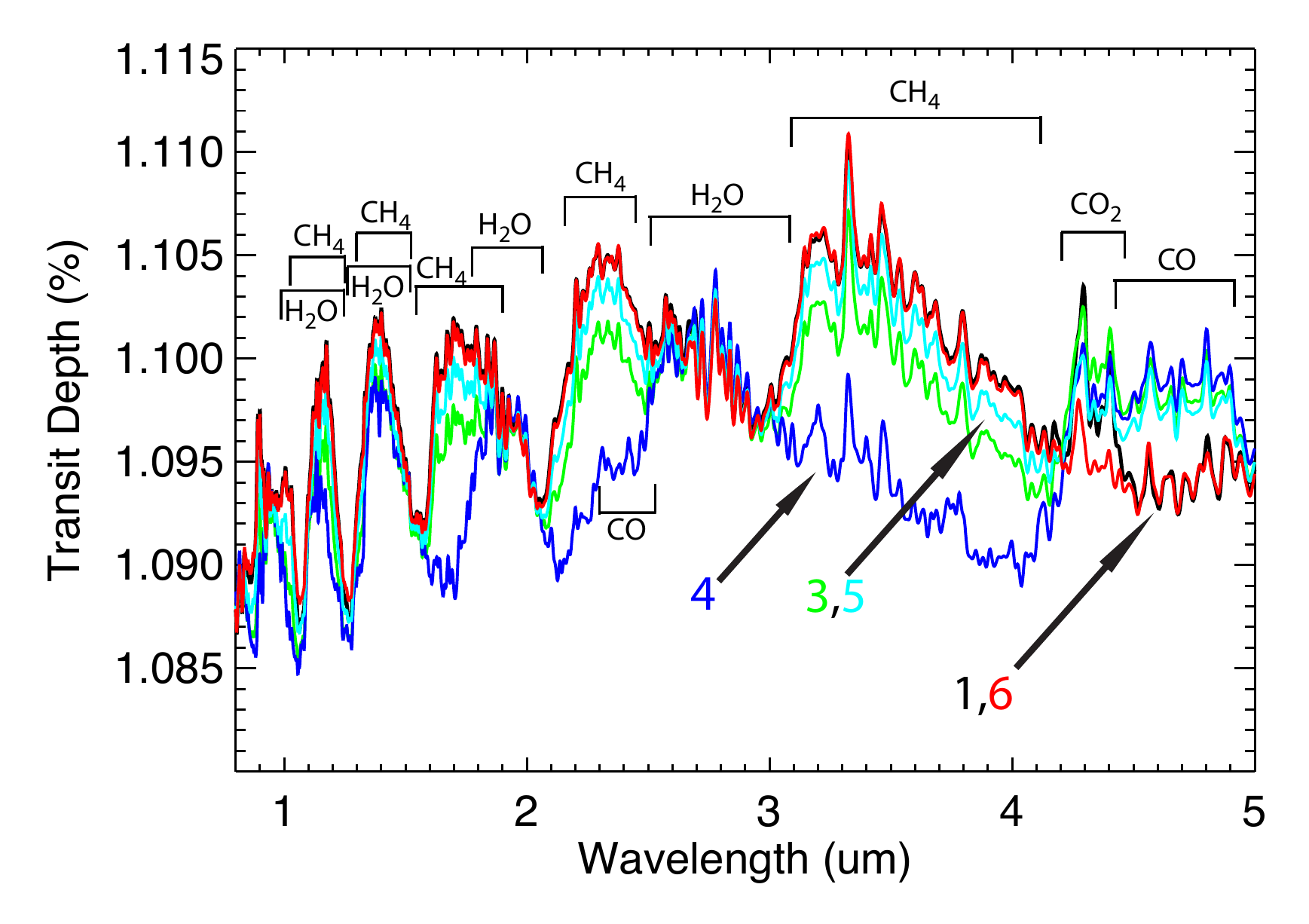}
\caption{The corresponding transmission spectra for the \emph{P--T} profiles and chemical abundance points from Figure \ref{profile}.  The main absorption features of H$_2$O, CO, CH$_4$, and CO$_2$ are labeled.  Transmission spectra that use the ``quenched'' chemical abundances from points 1, 3, 4, 5, and 6 are labeled with arrows.  Spectra are normalized to wavelengths where H$_2$O is the main absorber, to show the relative roles of CO and CH$_4$ in shaping spectra.  The transit models assume 1 \rj\ at a pressure of 1 kbar, a gravity of 25 m s$^{-2}$, and stellar radius of the Sun. 
\label{spectra1}}
\end{figure}

As one would expect, the spectra that use the quenched abundances, brought up to the visible atmosphere from the black points of Figure \ref{profile}, vary considerably as the abundances of CO and CH$_4$ vary by orders of magnitude.  In addition, the abundance of H$_2$O changes depending on whether CO is present as well.  We demonstrate this for 5 different models shown in Figure \ref{spectra1}.  For points to the ``right" of the CO/CH$_4$ equal-abundance curve, like 3, 5, and especially 4, the CO band is much stronger, and CH$_4$ weaker.  The spectra from points 1 and 6 are substantially similar, given their relatively positions in CO/CH$_4$ phase space.  The lack of monotonic behavior in the mixing ratio (and observability) of CH$_4$ as a function of the quench pressure was also pointed out for by \citet[see their Figure 2]{Mola19}, although they did not explore variations in the lower boundary condition, which is our focus here.

\begin{figure}[hbp!]
\includegraphics[clip,width=1.0\columnwidth]{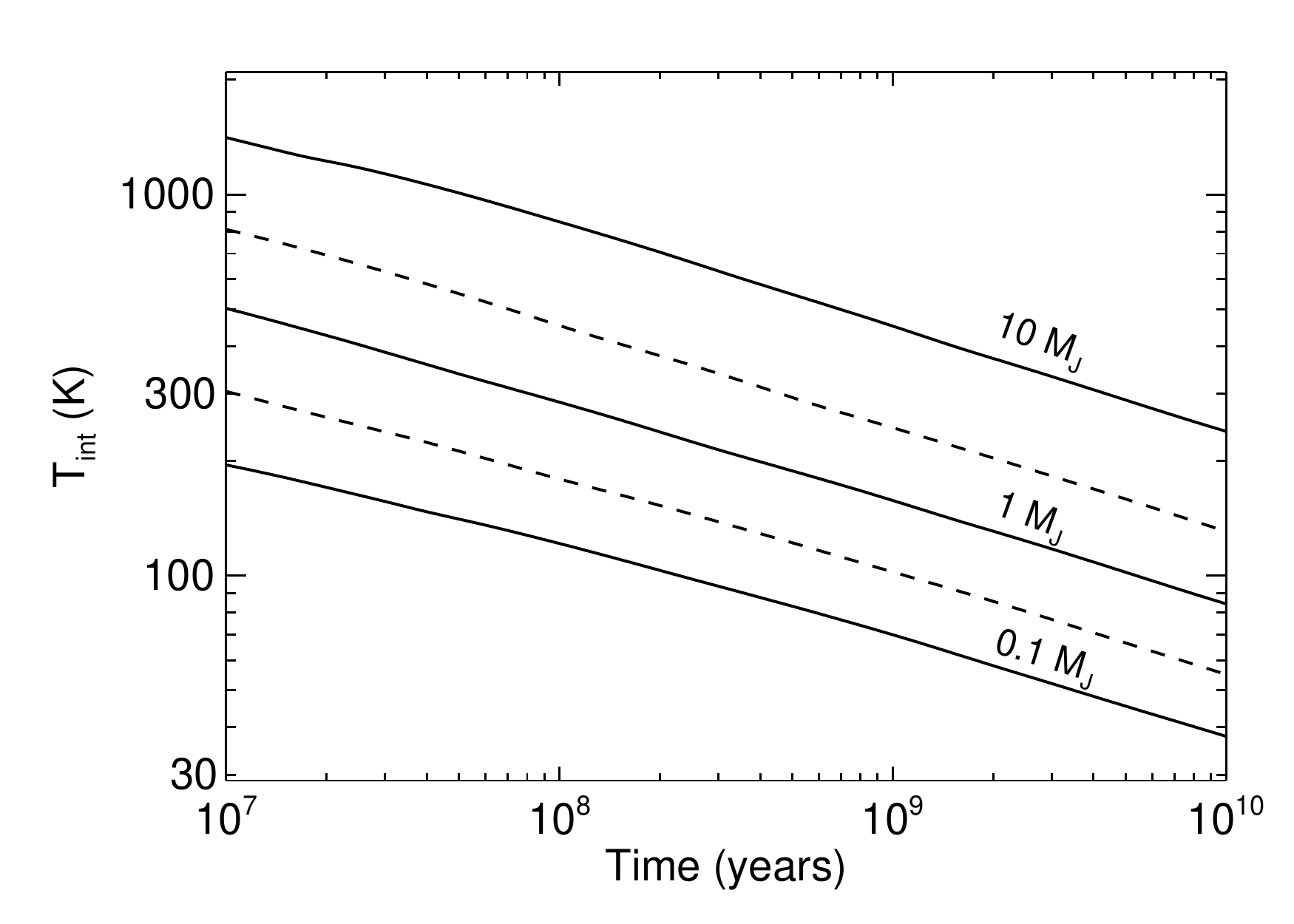}
\caption{Thermal evolution of giant planets at 0.1 AU from the Sun, after \citet{Fortney07a} and \citet{Thorngren16}.  Plotted are the intrinsic effective temperature, \tint, for models at 10, 3, 1, 0.3, and 0.1 \mj\ (32 \me), from top to bottom.  For reference, Jupiter today has \tint $=99$\,K.  A wide range of \tint\ values are possible at old ages, given a range of planetary masses, and a wide range of \tint\ values are possible at a given mass, over time.
\label{tracks}}
\end{figure}

\begin{figure}[hbp!]
\includegraphics[clip,width=1.0\columnwidth]{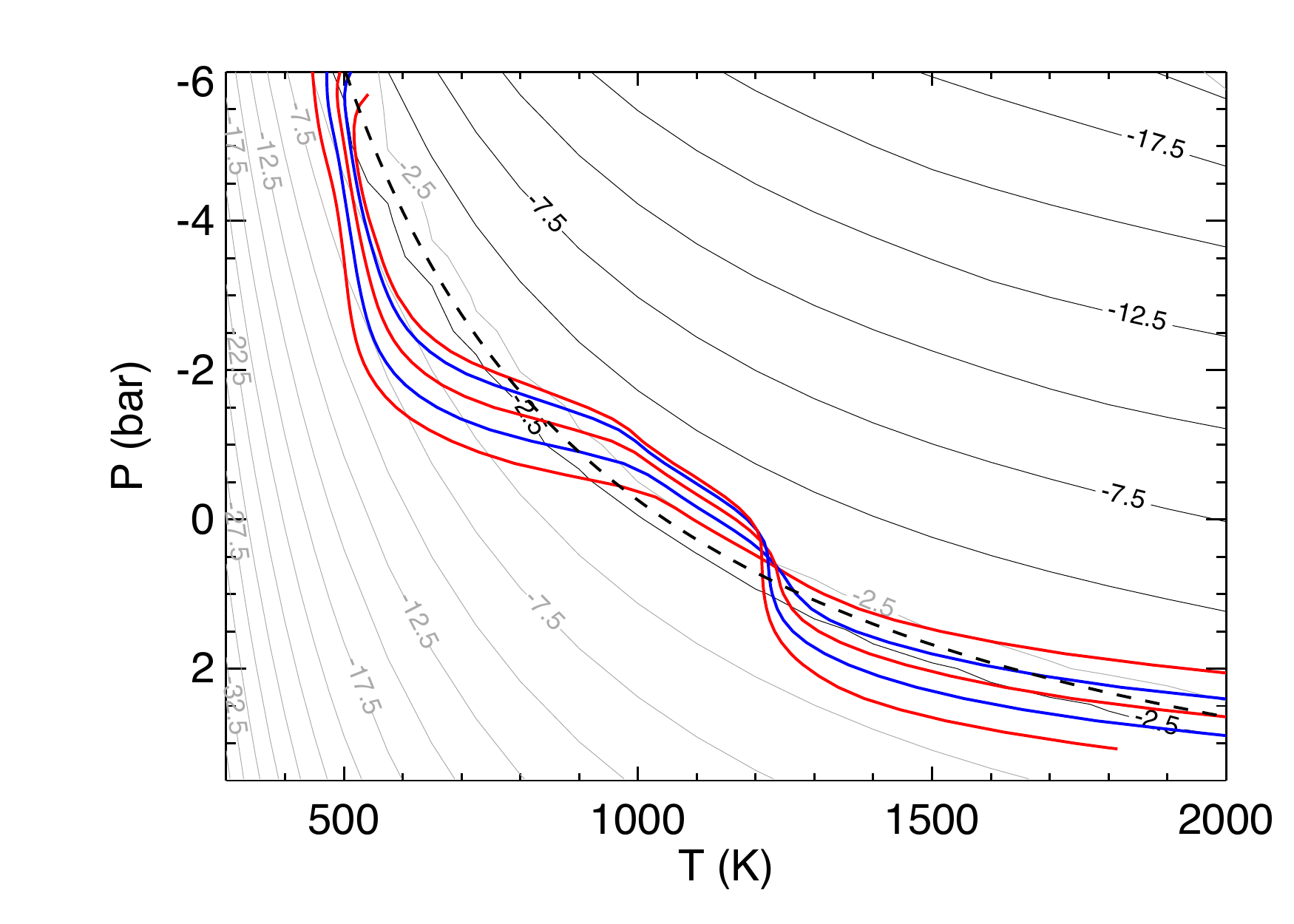}
\caption{Model pressure-temperature profiles (with \teq = 710 K) for a $10\times$ solar atmosphere at 0.15 AU from the Sun, this time based on thermal evolution models.  The five profiles (alternating red and blue) show five values of \tint, at 52, 77, 117, 182, and 333 K, as respective surface gravities $g$=5.8, 9.8, 24, 65, and 225 m s$^{-2}$.  Equal-abundance contours for CH$_4$ are shown in black, and light gray contours show the same for CO.  CO and CH$_4$ have an equal abundance at the dashed thick black curve.  These mixing ratio contours assume equilibrium chemistry.
\label{gravity}}
\end{figure}

Such a wide range of internal adiabats, for a given upper atmosphere, is quite possible due to the differences in cooling histories in giant planets.  It is by now widely appreciated that giant planets cool over time, most dramatically at young ages, and that more massive planets take longer to cool \citep{Marley96,Burrows97,Chabrier00}.  For reference, in Figure \ref{tracks} we plot cooling tracks for planets from 10 \mj\ to 0.1 \mj\ (32 \me) for ages from $10^7$ to $10^{10}$ years, using the models of \ct{Fortney07a} and \ct{Thorngren16}.  At an age of 3 Gyr, for instance, \tint\ values of 50 K to 350 K span the population.  Such model planets would in reality all have different surface gravities, which would then yield different \emph{P--T} profile shapes, even at the same orbital separation, as shown in Figure \ref{gravity}.  This plot is for the expected surface gravity for the five planet masses (at an age of 3 Gyr) shown in Figure \ref{tracks}.

Taken as a whole, these simple examples serve as motivation to explore a wider range of parameter space for H/He-dominated atmospheres.  The aim then is to show that a range of factors other than equilibrium temperature can have significant impacts, even \emph{dominant} impacts, on atmospheric abundances and spectra.  We also explore how non-equilibrium chemistry can serve as a tracer for understanding the deep temperature structure for these atmospheres, at pressures far below where one can probe directly.  After describing our methods in a bit more detail, we investigate these factors, first for well-known transiting Neptune-class planets GJ 436b, GJ 3470b, and WASP-107. After that we will explore carbon chemistry more generally, followed by nitrogen chemistry more generally, before our Discussion (with caveats), and Conclusions.

\section{Model Description}
\subsection{Atmospheric Structure and Spectra}
The model atmosphere methods used here have previously been extensively described in the literature.  We compute planet-wide average (``$4\pi$ re-radiation of absorbed stellar flux'') 1D radiative-convective equilibrium models using the model atmosphere code described in the papers of \citet{MM99}, \citet{Marley96}, \citet{Fortney05}, \citet{Fortney08a}, and the general review of \citet{Marley15}.  The radiative transfer methods are described in \citet{Mckay89}.  The model uses 90 layers, typically evenly spaced in log pressure from 1 microbar to 1300 bars.  The equilibrium chemical abundances follow the work of \citet{Lodders02},  \citet{Visscher06,Visscher10} and \citet{Visscher12}.  The opacity database is described in \citet{Lupu14} and \citet{Freedman14}.  Transmission spectra are calculated using the 1D code described in \citet{Morley17b}.

\subsection{Interiors and Tidal Heating}
As already mentioned, the giant planet thermal evolution models use the methods of \citet{Fortney07a} and \citet{Thorngren16}.  These thermal evolution calculations use an extensive grid of 1D non-gray solar-composition radiative-convective atmosphere models, which serve at the upper boundary condition.  The interior H/He equation of state is that of \citet{SCVH}.  We make the standard, typical assumption of a fully-convective H/He envelope, and these evolution models also have a 10 \me\ ice/rock core.

Tidal heating, to be investigated in a Section \ref{tides}, uses the extensive tidal evolution equations derived in \citet{Leconte10}.  We determine the tidal heating rate (in energy per second) with equation (13) in this work.  We will show that for some planets this tidal heating flux from the interior can be orders of magnitude higher than that calculated from normal secular cooling of the interior.

\subsection{Nonequilibrium Chemistry}
When treating non-equilibrium chemistry, an important topic in this paper, we make extensive use of the findings of \citet{Zahnle14}.  These authors provide quenching relations that are derived by fitting to the complete chemistry of a full ensemble of 1D kinetic chemistry models.  We use the standard ``quench pressure" formalism, where we assume chemical equilibrium where the chemical conversion time, \tchem, is shorter than the vertical mixing time, \tmix.  The local values of \tmix\ along a \emph{P--T} profile use the standard assumption that \tmix = $L^2 / K_{zz}$, where $L$ a length scale of interest, here assumed to be the local pressure scale height, $H$, and $K_{zz}$ is the vertical diffusion coefficient.  Other, potentially smaller values of $L$ could be used \citep{Smith98,Visscher11}, however, as we discuss below, uncertainties in $K_{zz}$ dwarf any uncertainty in $L$, so, following \citet{Zahnle14}, we make the simplest choice.

For these strongly irradiated planets, atmospheres can be radiative until depths of tens of bars, even beyond $\sim$1 kbar, depending on the the value of \tint.  The lower the value of \tint, the deeper the radiative zone, as shown in Figure \ref{profile}.  While in convective zones mixing length theory can be used as a guide to values of $K_{zz}$ \citep{Gierasch85}, in radiative regions no such readily usable theory exists, although it is generally expected that radiative regions will have orders of magnitude lower $K_{zz}$ values.

Some 3D circulation model simulations of hot Jupiters have attempted to gauge reasonable $K_{zz}$ values.  \citet{Parmentier13} suggested a fit to models of planet HD 209458b that yielded $K_{zz} = 5 \times 10^8 / \sqrt{P_{\rm bar}}$ cm$^2$ s$^{-1}$.  They suggest that cooler planets, like the ones treated here, should have slower vertical wind speeds and smaller values of $K_{zz}$.  More recent work has tried to estimate $K_{zz}$ from first-principles \cp{Zhang18a,Zhang18b,Menou19}.

The chemical kinetics literature for irradiated planets shows a range of $K_{zz}$ choices.  These include basing values tightly on 3D simulations, but more commonly, choosing a wide-range of constant-with-altitude $K_{zz}$ values, to bracket a reasonable parameter space.  It is this bracketing choice that we make here, as we aim to make the point that non-equilibrium chemistry must be important for a wide range of objects.  For calculations for particular planets of interest it may be worthwhile to generate \kzz\ predictions from GCM simulations.  We return to this point in Section \ref{discussion}.  Followup work that couples planetary temperature structures with detailed predictions of \kzz\ profiles \citep{Zhang18a,Zhang18b,Menou19}, to predict atmospheric abundances, would be important and fruitful work.

Before exploring a wide range of planets, we first investigate how our models can be used to understand the atmospheric abundances of three (relatively) well-studied Neptune-class transiting planets, which have been the targets of many observations with \emph{Spitzer} and \emph{Hubble}.

\section{The Atmospheres of Three Neptune-Class Planets: GJ 436b, GJ 3470b, WASP-107b} \label{tides}
Our first foray into why \teq\ is not enough will be for the Neptune-class exoplanets, \gjf, \gjt, and \wasp.  These three planets have been the targets of extensive observational campaigns, in particular for \gjf, as it was the first transiting Neptune-class planet found \cp{Gillon07}. The work on emission and transmission observations and their interpretation for this planet is large and difficult to concisely summarize.  A recent review can be found in \citet{Morley17a}.  The most significant finding, going back to \citet{Stevenson10}, is the suggestion that the planet's atmosphere is far out of chemical equilibrium, with little CH$_4$ absorption and a likely high abundance of CO and/or CO$_2$.  An upper limit on the CH$_4$ abundance is published in \citet{Moses13}.

More recently, \citet{Benneke19} found that a joint retrieval of the emission and transmission data for \gjt\ points to a somewhat similar conclusion, with a lack of CH$_4$ seen.  And a transmission spectrum of \wasp\ by \citet{Kreidberg18} finds no sign of CH$_4$ in the near infrared.  For both planets, these papers include CH$_4$ abundance upper limits.

While these three planets have masses and radii that differ by a factor of around 2, they share some interesting similarities.  Perhaps most strikingly, they have \teq\ values that all within $\sim$~100 K of each other.  This \emph{may} suggest that the planets could have similar atmospheric properties.  Another, perhaps surprisingly fact, is that all three planets are on eccentric orbits.  Most important to our current discussion is that we find all three planets are currently undergoing significant eccentricity damping today.

\begin{figure}[htp]
\includegraphics[clip,width=1.0\columnwidth]{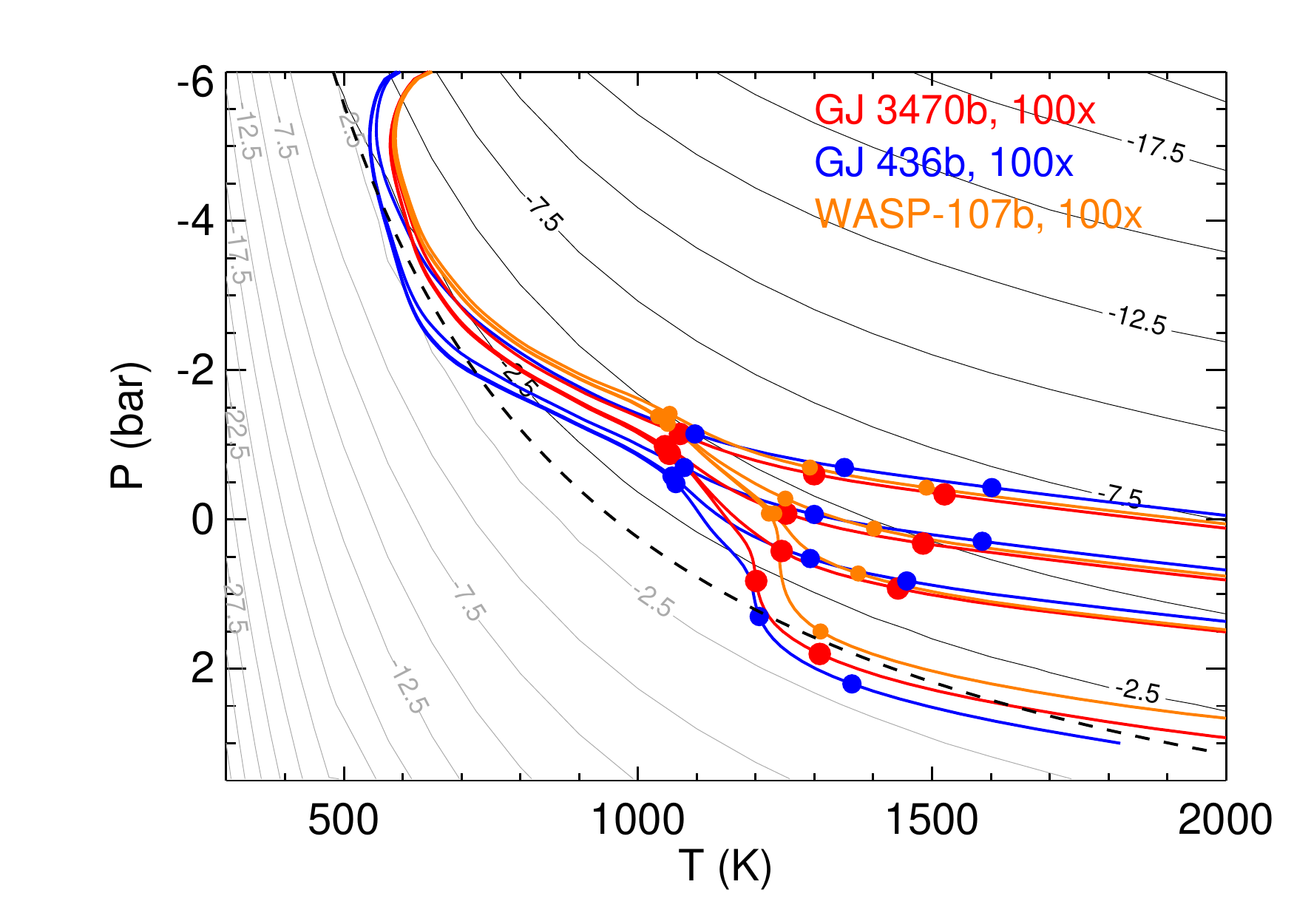}
\caption{Atmospheric \emph{P--T} profiles for planets GJ 436b, GJ 3470b, and WASP-107b all at $100 \times$ solar abundances.  The light and dark gray equal-abundance curves are similar to those in Figure \ref{profile}, although here we plot $100 \times$ solar.  For each planet, 4 interior adiabats are shown, for the case of no tidal heating (coolest), and $Q=10^6$, $10^5$ and $10^4$, from cooler to warmer.  The sets of solid dots show the quench pressure for log $K_{zz} =$ 4, 8, 12, where larger $K_{zz}$ values probe deeper.
\label{3profiles}}
\end{figure}

Figure \ref{3profiles} shows model \emph{P--T} profiles for all three planets, with GJ 436b in blue, GJ 3470b in red, and WASP-107b in orange.  For simplicity, all are at $100 \times$ solar, a value similar to the carbon abundance inferred for Uranus and Neptune.  We note that retrieval work for GJ 436b \citep{Morley17a} suggests a metallicity higher than this value, retrievals for GJ 3470b suggest a metallicity lower than this \citep{Benneke19}, and preliminary structure models (that did not take into account tidal heating) for WASP-107b also suggested a lower metallicity \citep{Kreidberg18}.  Our aim here is not to find best fits for the spectra of each planet, but to suggest that tidal heating in the interior plays a large role in altering atmospheric abundances.  We therefore feel that a simple, but plausible metallicity, can serve as an illustrative example.

A cursory glance shows that all 3 planets reside in a \emph{remarkably similar} \emph{P--T} space.  For these planets 4 adiabats are shown.  First we will examine the coolest adiabats (lowest specific entropy), which are for models with no tidal heating (\tint\ $=60 $K), and then 3 warmer adiabats that assume log $Q$ = 6, 5, and 4, from colder to hotter, as a lower $Q$ means more tidal heating \citep{Leconte10}.  Tidal heating for these planets has a dramatic effect, warming the interior by hundreds to \emph{thousands} of K at a given pressure.

\begin{figure}[htp]
\includegraphics[clip,width=1.0\columnwidth]{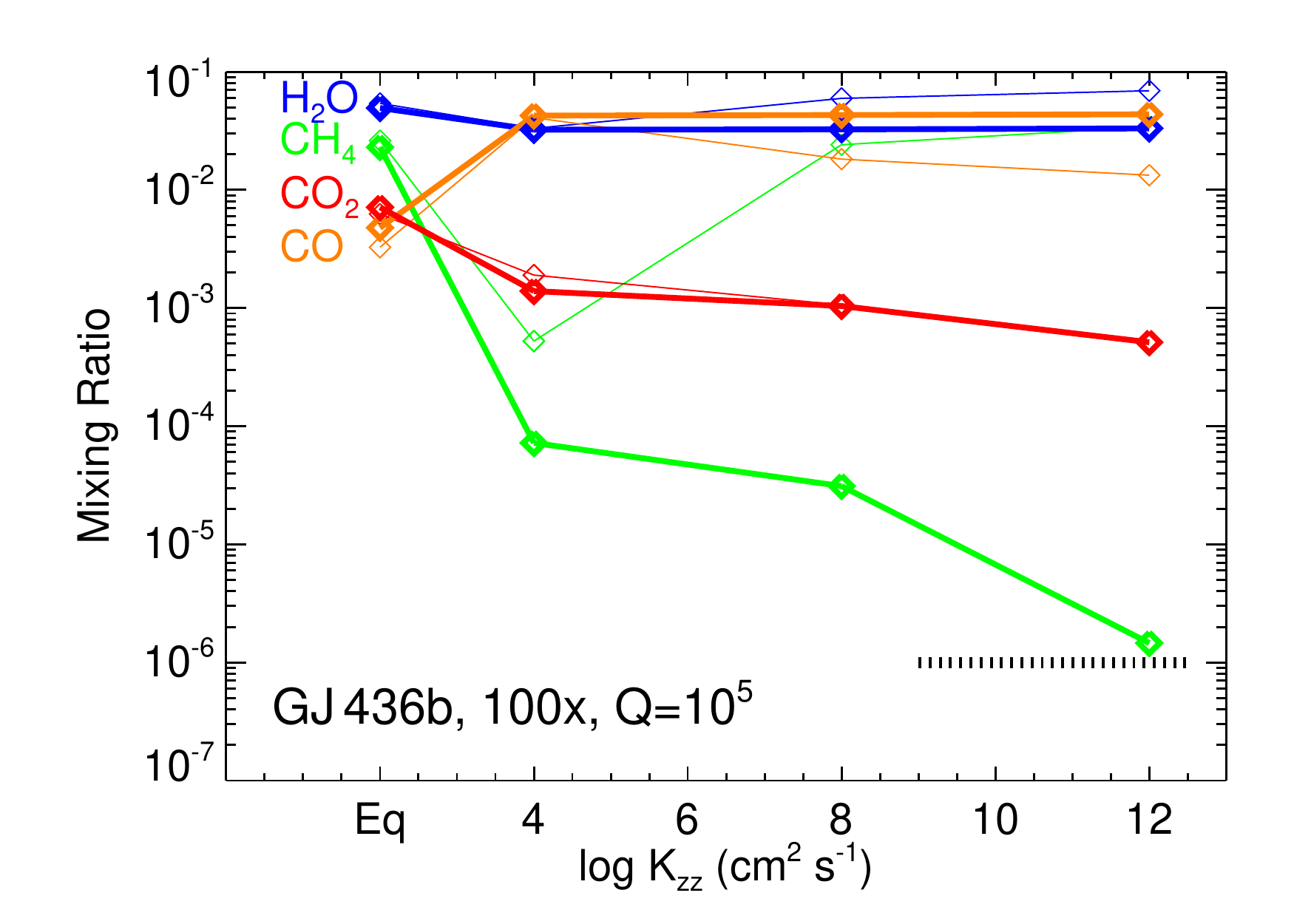}
\includegraphics[clip,width=1.0\columnwidth]{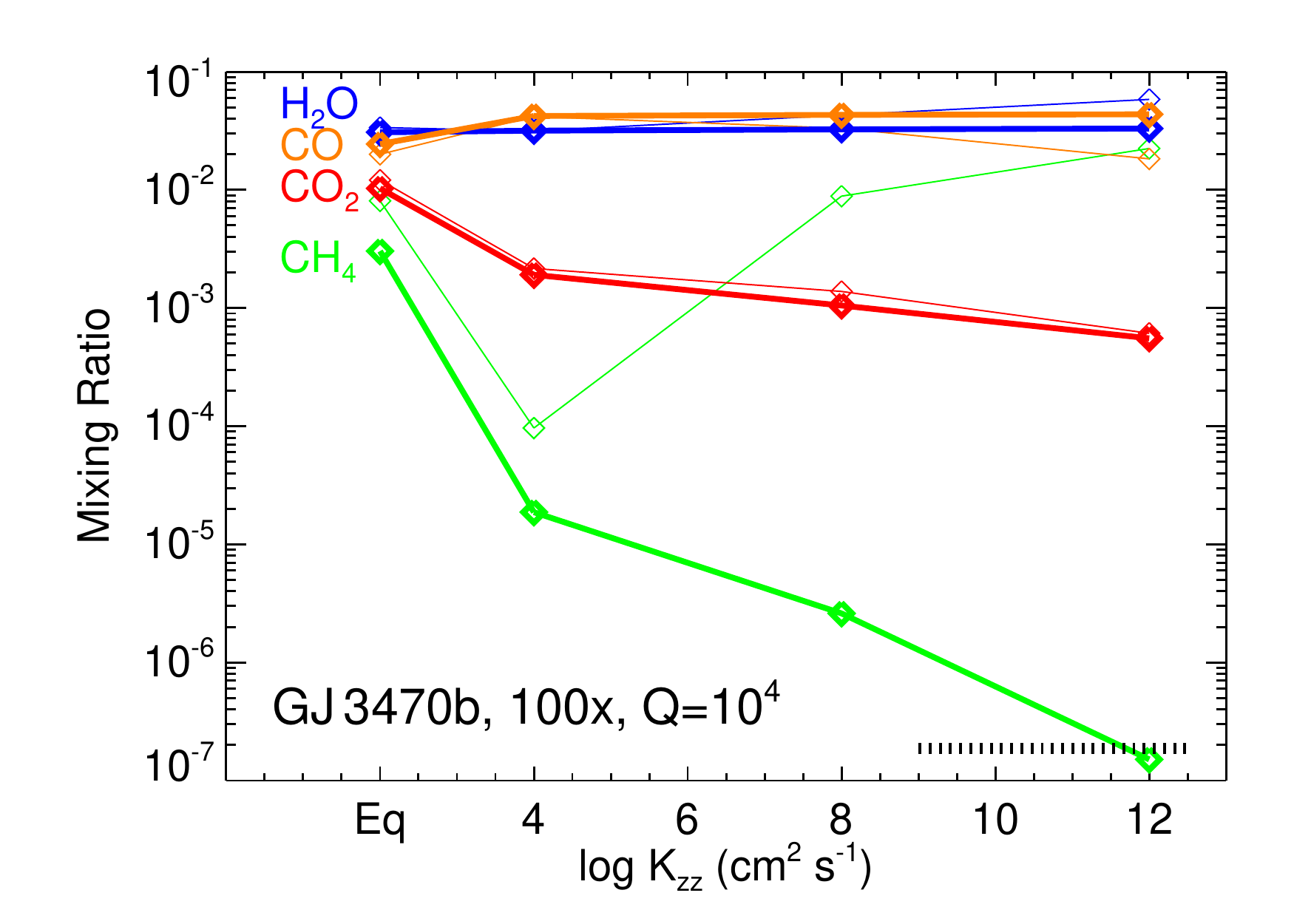}
\includegraphics[clip,width=1.0\columnwidth]{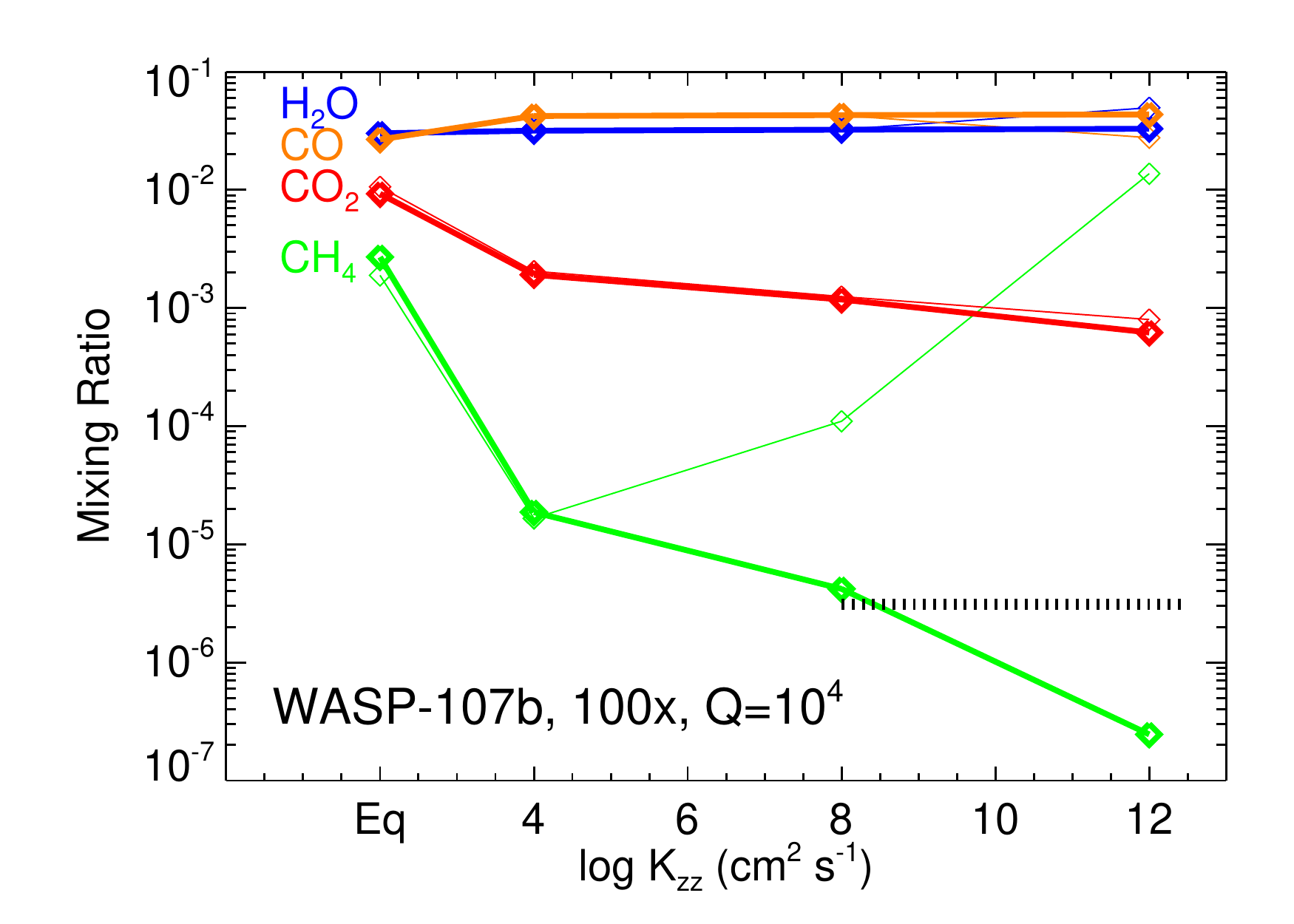}
\caption{Top: Chemical abundances at 1 mbar for 3 models of GJ 436b.  H$_2$O is blue, CO is orange, CO$_2$ is red, and CH$_4$ is green.  Plotted are abundances for equilibrium chemistry, and log $K_{zz}$= 4, 8, and 12. Thin lines show no tidal heating, while thick lines use $Q=10^5$.  With tidal heating, the higher the $K_{zz}$, the higher the CO/CH$_4$ ratio.  The dashed black line shows the CH$_4$ mixing ratio upper limit.  Middle: A very similar plot for GJ 3470b, again showing how nonequilibrium chemistry and tidal heating enhance the CO/CH$_4$ ratio, but with $Q=10^4$.  Bottom:  Another similar plot for WASP-107b, with $Q=10^4$.  Tidal heating and high \kzz\ can plausibly explain all observations.  \label{3chemplots}}
\end{figure}

All three planets have three sets of solid dots on their profiles that show the quench pressure level for log \kzz\ = 4, 8, and 12 cm$^2$ s$^{-1}$\footnote{log \kzz\ $\sim10.5$ is the maximum allowed from mixing length theory, for GJ 3470b and WASP-107b, for the hottest interior profiles shown, per equation 4 from \citet{Zahnle14}.}.  For the quench pressure for log \kzz = 4, very sluggish mixing, tidal heating has a modest impact in shifting the expected chemical abundances to CO-richer and CH$_4$-poorer territory, compared to, say, equilibrium chemistry at 1 mbar. However, for the depths probed at log \kzz = 8 and 12, the atmosphere models are significantly warmer, and draw from a region of much higher CO and lower CH$_4$ if heating is present.  We can explore and quantify this effect for a subset of models, which are shown in Figure \ref{3chemplots}, where each planet has its own panel.  Abundances at 1 mbar are plotted for equilibrium chemistry and log \kzz\ = 4, 8, and 12.  Thin lines are for no tidal heating, while thick lines include tidal heating, with $Q=10^4$ -- a reasonable estimate for Neptune \cp{Zhang08} -- for GJ 3470b and WASP-107b, and $Q=10^5$ for GJ 436b, based on a fit to the planet's thermal emission spectrum \citep{Morley17a}.  At our assumed $100\times$ abundances with equilibrium chemistry, for all three planets CH$_4$ would be expected to be abundant, and even the dominant carbon carrier in GJ 436b and WASP-107b.  The retrieved 1$\sigma$ CH$_4$ upper limits, from free retrievals from all three atmospheres \citep{Moses13,Kreidberg18,Benneke19}, are shown as dashed black lines.

There are two main effects to be seen in Figure \ref{3chemplots}.  First in the large change in abundances for CH$_4$ -- falling off dramatically, and CO -- increasing, but more modestly, just in going from equilibrium chemistry to log \kzz = 4.  Another striking effect is the divergence in the behavior of the CH$_4$ abundance at log \kzz = 8 and 12, between the no tidal heating model (thin lines) and the model with tidal heating.  Based on the \emph{P--T} profiles in Figure \ref{3profiles} we can see that no-heating models bring up CH$_4$-rich gas, while the tidal heating models bring up CH$_4$-poor gas.  This is a dramatic effect in all three planets.  Large \kzz\ values, driven by strong convection caused by ongoing tidal dissipation, can drive the CH$_4$ abundance to low values, in the range constrained by observations to date.

This strongly suggests that nonequilibrium chemistry and tidal heating conspire to drive the atmospheric abundances far from simple expectations.  We should of course be a bit wary about treating the three planets as \emph{carbon copies} however.  With no theory to guide the strength of tidal heating, $Q$ for the planets could be quite different for all three.  The expected mass fraction of H/He in WASP-107b is far larger than for GJ 3470b, for instance.  Similarly, with little theory to guide vertical mixing strength, this could also be quite different among the planets, as they have quite different surface gravities.  Additionally, they have been modeled with relatively simple chemical abundances ($100 \times$ solar, with a solar C/O ratio), and the actual planets could readily have more complex, and different, base elemental abundances.  Of note, the planet WASP-80b, about $100-150$\,K warmer than this trio, but on a circular orbit \citep{Triaud15}, has a \emph{Spitzer} IRAC 3.6/4.5 $\mu$m ratio in thermal emission that is similar to early T-dwarfs.  \citet{Triaud15} suggest this IRAC color could potentially be due to some CH$_4$ absorption in the planet's atmosphere, which seems quite viable, as we describe in the next section.

As \citet{Morley17a} suggested for GJ 436b, a direct sign of tidal heating would be a high thermal flux from the planet's interior, which could be observed via a secondary eclipse spectrum or thermal emission phase curve.  Future observations with \emph{JWST}, including those where tidal heating are not at play, may allow for a coupled understanding of atmospheric abundances, temperature structure at a variety of depths, vertical mixing speed, and tidal heating.  These three planets, all in a similar \emph{P--T} space, motivate a wider investigation.

\section{The Phase Space of Chemical Transitions} \label{transitions}
In the face of vertical mixing altering chemical abundances, mixing ratios in the visible atmosphere are tied to atmospheric temperatures at depth, as described in the previous section. This complicates the goal of deriving a straightforward understanding of chemical transitions. We aim to show that, even at a given metallicity and \kzz, this transition will depend on the cooling history (hence, mass and age) of any planet.  We refer back to Figure \ref{tracks} which showed models of the thermal evolution of giant planets. These model planets are all at 0.1 AU from the Sun, but these cooling tracks would be correct, to within several K, at closer or farther orbital distance \cp{Fortney07a}. Therefor, we can investigate, at a fixed value of \tint, how changing incident flux (hence, \teq) does or does not lead to changes in chemical abundances in the visible atmosphere.  We first explore carbon chemistry. 

\subsection{CO-CH$_4$ Transitions} \label{carbon}
In Section \ref{tides} we examined the CO-CH$_4$ boundary for specific tidally-heated Neptune-class planets.  Objects with tidal heating are special cases, but certainly will be common enough that they cannot simply be ignored, when looking at general trends.  But here we can examine the general trends in the absence of tidal heating, for a range of planet masses and ages.  As we will see, the range of cooling histories, and lack of clarity with how vertical mixing will change with planet mass, can lead to important complexities.

\subsubsection{Effects of \teq\ and Vertical Mixing}
\begin{figure}[htp]
\includegraphics[clip,width=1.0\columnwidth]{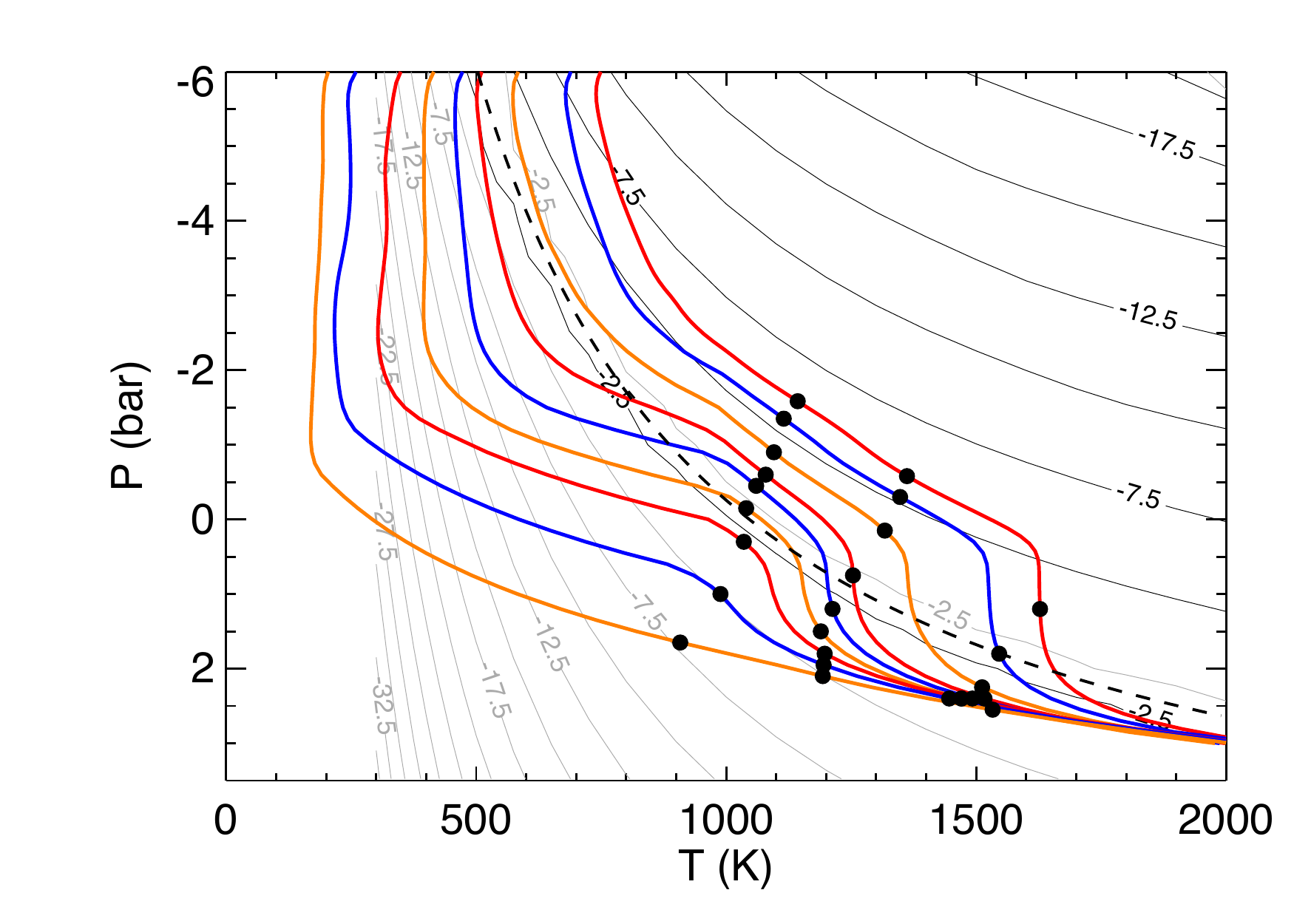}
\caption{Atmospheric \emph{P--T} profiles for old, Saturn-like planets (\tint$=75$K, $g=10$m s$^{-2}$, assuming $10\times$ metallicity.  The models are a 9 incident flux levels, at 0.06, 0.07, 0.1, 0.15, 0.2, 0.3, 0.5, 1, 2 AU from the Sun.  Three sets of black dots show the depth of vertical mixing with log \kzz\ of 4, 8, and 11 cm$^2$ s$^{-1}$.  At higher pressures, note that the spread between all profiles is lessened, both in temperature, and in reference to the CH$_4$ (black) and CO (grey) abundance curves.)
\label{saturnprofs}}
\end{figure}

We first examine the general case of a Saturn-like exoplanet as a function of distance from a Sunlike star.  Here we have chosen a $10\times$ solar atmosphere, surface gravity of 10 m s$^{-2}$, and \tint $=75$ K, representative of a several gigayear-old Saturn-mass exoplanet.  We choose this as our ``base planet'' since these kinds of giant planets would be excellent targets for atmospheric characterization via transmission.  Atmospheric \emph{P--T} profiles are shown in Figure \ref{saturnprofs}, for planets from 0.06 AU to 2 AU.  The three sets of black dots show quench pressures corresponding to log \kzz\ values of 4, 8, and 11.  Most importantly, at lower pressures, the atmospheres diverge quiet widely, owing to the factor of $\sim$1100 difference in incident flux across these models.

As one looks deeper it is apparent that profiles modestly converge as the pressure increases, followed by a dramatic ``squeezing together" as the planets fall on nearly identical adiabats.  This is a generic behavior for $g/$\tint\ pairs, and one could make a plot like this for any Jupiter-like planet, super-Jupiter, or sub-Saturn.  Why this behavior occurs requires some discussion.  To our knowledge this effect was first noted in Figure 3 of \citet{Fortney07a}, who described the effects of these ``bunched up" deep profiles on the mass-radius relation for warm transiting giant planets, but they did not identify a cause for the similarity of the deep temperatures.

A study of the gray analytic temperature profiles of \citet{Guillot10} suggests, via their Equation (29), a relation between the temperature ($T$) and optical depth $\tau$ that is a function of only three quantities:  the irradiation temperature (which is directly related to \teq), \tint, and  $\gamma$, the ratio of the visible to thermal opacities.  If $\gamma$ is relatively constant, and at a given \tint\ value, decreasing \teq\ cools the entire atmosphere at every $\tau$, including the deep region that here transitions to an adiabat.  However, if $\gamma$ were to dramatically decrease with decreasing \teq, the deep $T -- \tau$ profile (analogous to our deep \emph{T--P} profile) could remain nearly constant at depth with an upper atmosphere that was colder with decreasing \teq.  Indeed, Figure 5 of \citet{Freedman14} shows a factor $\sim$60 falloff in $\gamma$ from $\sim$1400-700\,K, due to the loss of alkali metals Na and K from the vapor phase, with $\gamma$ relatively constant at hotter and colder temperatures.  This 700-1400 \,K temperature range corresponds reasonably well to what is seen in our Figure \ref{saturnprofs} and ``middle region" of Figure 3 of \citet{Fortney07a}.  Therefore, we suggest that this change in visible opacity is the dominant physical effect the keeps the deep atmosphere temperatures relatively constant across this \teq\ range.  However, additional work on this point is surely needed.

Of particular interest is that the coldest profiles are mostly in the CH$_4$-dominant region at lower pressures, but along the atmospheric adiabat, as one reaches hotter layers, one finds gradually more CO.  This is the ``typical" case for brown dwarfs \citep{Saumon03,Phillips20} and for Jupiter as well \citep{Prinn77,Lodders02}.  However, for the hottest models, this typical trend is reversed, and when one probes quite deeply, one reaches more CH$_4$-rich gas, in particular at $P> 1$ bar, where the isothermal regions are reached.

\begin{figure}[htp]
\includegraphics[clip,width=1.0\columnwidth]{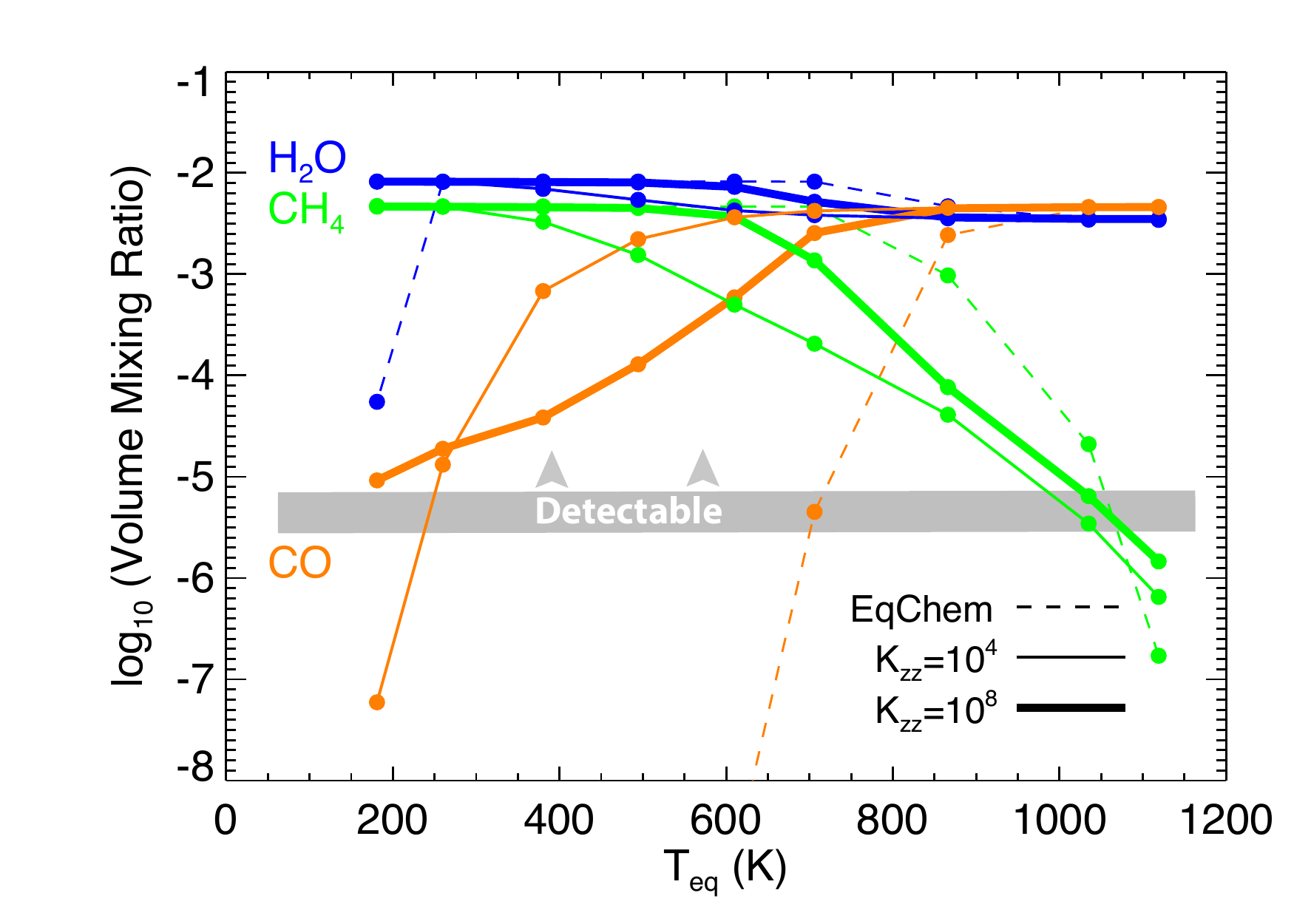}
\caption{The 9 \emph{P--T} profiles from Figure \ref{saturnprofs} are plotted at 9 \teq\ values across the x-axis, with chemical abundances along the y-axis.  ``EqChem" gives the chemical equilibrium abundances at 1 mbar (dashed), while log \kzz\ $=4$ and 8 are shown as thin solid and thick solid, respectively.  In equilibrium, at \teq\ $<800$ K, the CO mixing ratio falls off precipitously, while for log \kzz\ $=4$ this falloff is delayed until $\sim$~500 K cooler.  At log \kzz\ $=8$ the weakening of CO is also delayed and the change in CO abundance with \teq\ is much ``shallower."  The corresponding increases in CH$_4$ abundance with lower \teq\ is again ``shallower" for non-equilibrium chemistry.  The loss of H$_2$O in the coolest (equilibrium) model is due to loss of water vapor into water clouds.
\label{saturnchem}}
\end{figure}

We can examine how atmospheric abundances are affected by making plots of volume mixing ratio as a function of planetary \teq.  Such a plot is shown in Figure \ref{saturnchem}, and includes all the profiles shown in Figure \ref{saturnprofs}.  The mixing ratios at 1 mbar for H$_2$O, CO, and CH$_4$ are plotted, for equilibrium chemistry and for log \kzz\ of 4 and 8.  In the equilibrium chemistry case (dashed curves), the changeover from CO-dominant to CH$_4$ dominant is at about \teq $=850$ K.  As one goes cooler, this also leads to an increase in the H$_2$O abundance, as oxygen is liberated from CO (and CO$_2$).

If we include quite sluggish vertical mixing, with log \kzz $=4$ (thin solid line), this boundary shifts dramatically left, to a much lower \teq\ value of only 475 K.  The slopes of the CH$_4$ and CO curves, vs. \teq, are both quite shallow compared to the equilibrium chemistry case and one might readily expect both molecules to be seen from $\sim$800 to 200 K.  Of course how ``detectable" a molecule is depends strongly on the wavelength being investigated, the spectral resolution, and the impact on other opacity sources, like clouds.  Given the non-detections of CH$_4$ with \emph{HST} at mixing ratios of $\sim 10^{-6}$ in the Neptune-class planets (See Section \ref{tides}), here we suggest $\sim 10^{-5.5}$.  However, he 3.3 and 7.8 $\mu$m bands of CH$_4$ and 4.5 $\mu$m band of CO are strong and could likely yield detections at lower mixing ratios, in particular at high spectral resolution.

Interestingly, a look back to Figure \ref{saturnprofs} might suggest that log \kzz $=8$ case might be a bit less extreme in altering abundances, even though we are mixing up from even hotter layers.  The modest pinching together of the \emph{P--T} profiles yields a behavior in Figure \ref{saturnchem} (solid line) that is intermediate between the two previous behaviors, with a crossover \teq\ of 680 K.  Both CO and CH$_4$ may be seen from \teq $\sim$~900 to 400 K.  The upshot here is that the value of \kzz\ in these atmospheres, and its depth dependence, which is currently unknown, will have a significant effect on the atmospheric abundances as a function of \teq, and a wide range of behavior is expected.  As discussed later, given that \kzz\ is unlikely to be constant with altitude, more realistic mixing further complicates this picture.

\subsubsection{Effects of Planet Mass at a Given Age} \label{mass1}
In the previous section we examined one particular planet, a Saturn-like object at different distances from the Sun.  However, we have already discussed in some detail in the Introduction that planets of different masses are expected to have quite different cooling histories (Figure \ref{tracks}).  

\begin{figure}[htp]
\includegraphics[clip,width=1.0\columnwidth]{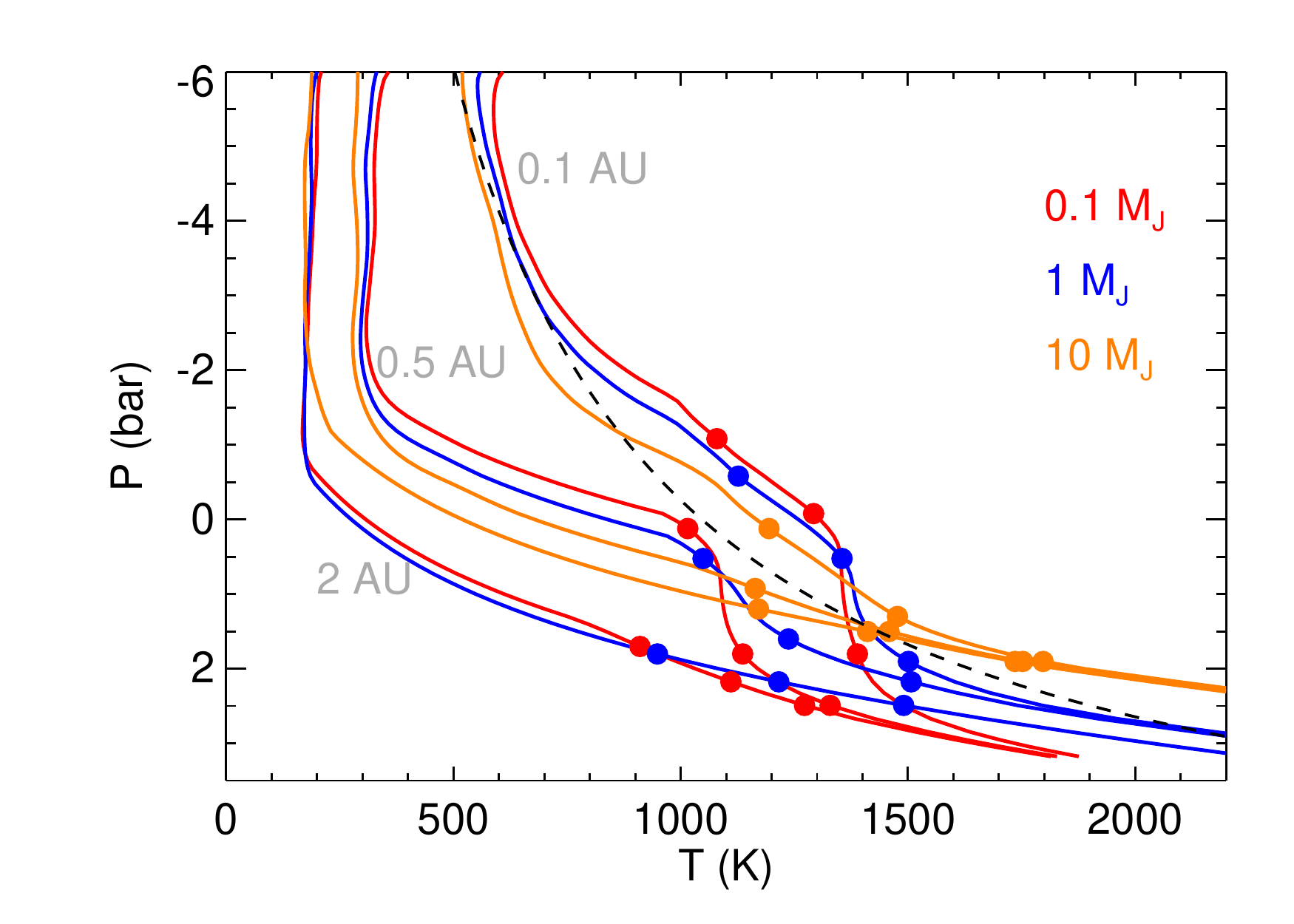}
\caption{Atmospheric \emph{P--T} profiles for 3-Gyr-old planets at 0.1 (red), 1 (blue), and 10 (orange) \mj, at 10$\times$ solar. The CO/CH$_4$ equal-abundance curve is in dashed black. The models are at 0.1, 0.5, and 2 AU from the Sun. The color-coded dots show the quench pressure for log \kzz\ $=4$, 8, and 11.  Higher gravity models have higher pressure photospheres, but also have hotter interiors, which causes significant crossing of profiles. The much larger scale heights for the low gravity models means greater physical distances for mixing, and hence, lower quench pressures.
\label{massprofs}}
\end{figure}

We can begin to address the question of planet mass with three disparate planet examples, with planets of 10 \mj\ (a super-Jupiter), 1 \mj\, and 0.1 \mj\ (32 \me, a super-Neptune). For now we limit ourselves to the same 10$\times$ atmospheric metallicity, so as to not change too many parameters at once.  Similar to Figure \ref{saturnprofs} above, we have computed a range of atmospheric \emph{P--T} profiles for these 3 planets, at different distances from the Sun, assuming an age of 3 Gyr and the \tint\ values from Figure \ref{tracks}.  These profiles are shown in Figure \ref{massprofs}.  For clarity, profiles are only shown at three distances, 0.1, 0.5, and 2 AU.  Along each profile, colored dots, from lower to higher pressure, show log \kzz\ of 4, 8, and 11, respectively.  The more massive the planet, the higher the surface gravity, and the higher pressure at a given temperature, in the outer atmosphere. This, however, is reversed in the deep atmosphere and interior as the higher mass planets take longer to cool, so they have a higher \tint\ (333 K, 117 K, and 52 K, respectively for the 10, 1, 0.1 \mj\ models) and ``hotter'' (higher specific entropy) interior adiabat.  The much larger scale heights for the low gravity models means greater physical distances for mixing, thus longer mixing times for a fixed \kzz, and hence, lower quench pressures.

\begin{figure}[htp]
\includegraphics[clip,width=1.0\columnwidth]{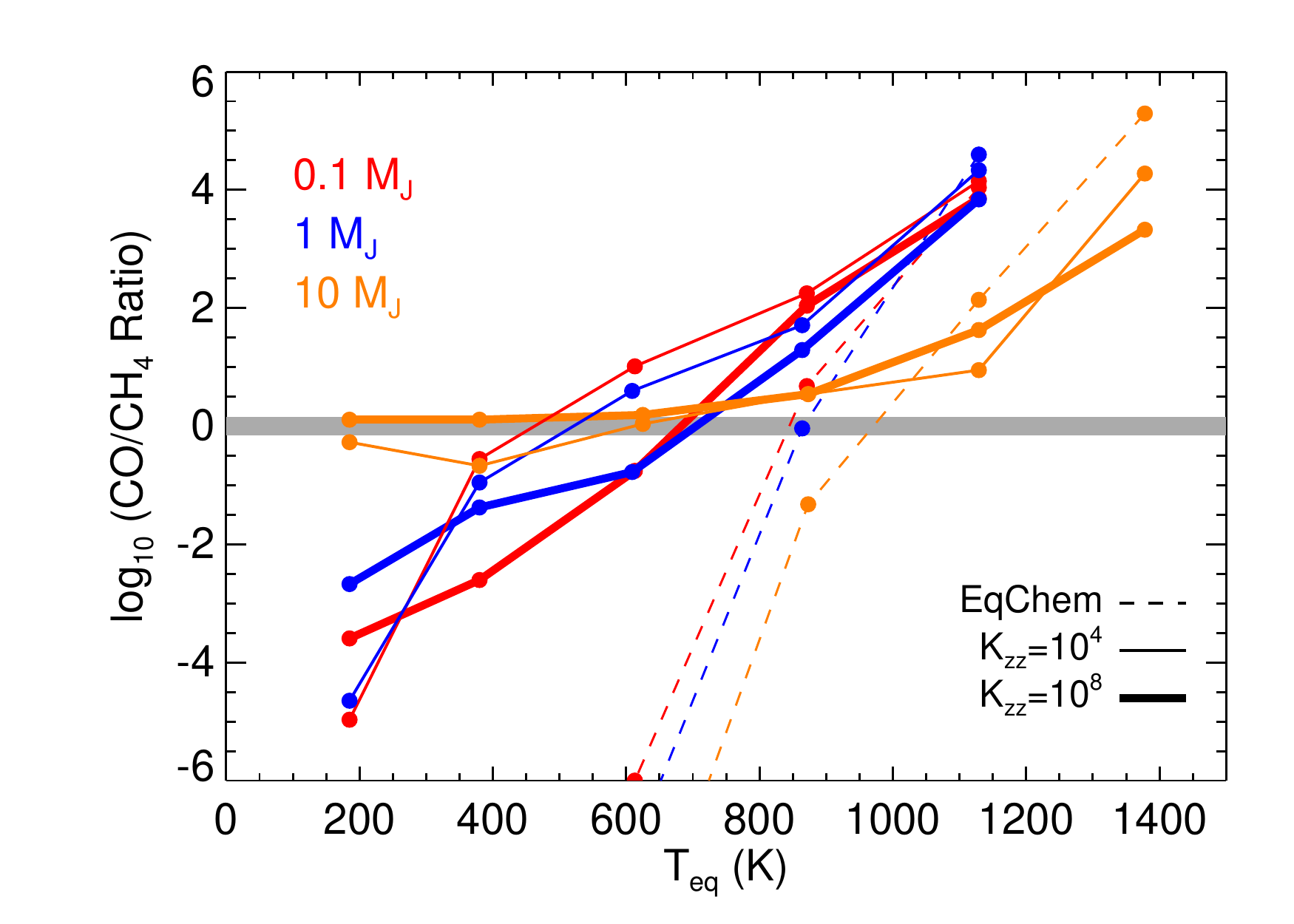}
\caption{The log of the CO/CH$_4$ ratio for 5 values of \teq\ for 0.1, 1, and 10 \mj\ model planets, where a subset of the profiles are shown in Figure \ref{massprofs}.  In equilibrium (at 1 mbar), the transition \teq\ for CO/CH$_4$=1 (log=0, shaded grey) is at $\sim$ 800, 950, and 1150 K, from low mass to high mass.  As expected, vertical mixing lessens the slopes of these curves, and pushes the transition \teq\ lower for the 0.1 and 1 \mj\ models. The 10 \mj\ model quenches from CH$_4$-richer gas, at high \teq, which yields the opposite behavior.  For all three model planets, CO and CH$_4$ exist together in detectable amounts for a wide swath of \teq\ values.
\label{masschem}}
\end{figure}

What we are particularly interested in here is how the role of surface gravity and cooling history work to dramatically change the ratio of CO/CH$_4$ in these atmospheres.  We address this scenario in Figure \ref{masschem}.  This abundance ratio is plotted vs.~planetary \teq\ and we will first examine the abundances for equilibrium chemistry at 1 mbar.  The ``transition" \teq\ value is 950 K at 10 \mj, and 850 K at 1 and 0.1 \mj.  With sluggish vertical mixing (log \kzz $=4$), the story becomes more complex, however.  The 10 \mj\ planet has a relatively hot interior adiabat, which is essentially the same for all values of \teq, as seen in orange in Figure \ref{massprofs}.  For such a large value of \tint, the smaller values of \teq\ becomes essentially irrelevant.  For the lower mass planets, the transition \teq\ is much lower than in the equilibrium case, reaching 500 K. For more vigorous mixing (log \kzz $=8$), more CH$_4$-rich gas is brought up, leading to a \emph{hotter} transition temperature, at 700 K.

\subsubsection{Effects of Planet Age at a Given Mass}  \label{cooling1}
Up until this point, we have examined ``old'' planetary systems that to date make up the vast majority of the transiting population.  However, studying younger transiting planets to better understanding evolutionary histories is extremely important.  First, this would yield connections to the directly imaged self-luminous planets, which are predominantly young \citep{Bowler16}.  Second, understanding atmospheric abundances as a function of planet age would give us new insight into planetary thermal evolution.  Third, since parent stars are much more active when they are young, high XUV fluxes for young systems could drive quite interesting photochemistry.

\begin{figure}[htp]
\includegraphics[clip,width=1.0\columnwidth]{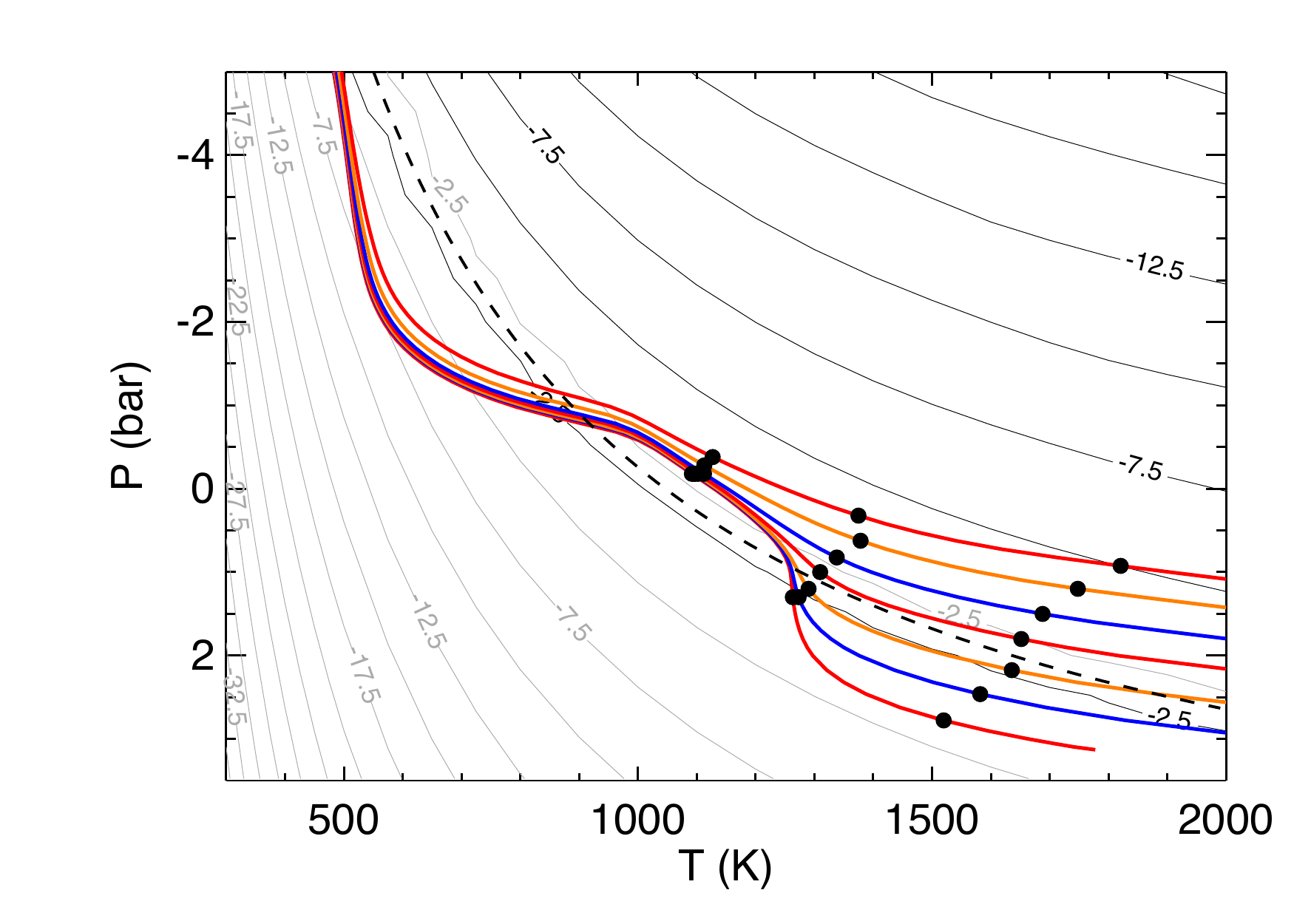}
\caption{Atmospheric \emph{P--T} profiles for a 1 \mj\ planet at 0.15 AU from the Sun, assuming $3\times$ solar metallicity.  Seven ages, every half dex from 10 Myr to 10 Gyr, with seven values of \tint\ (501, 383, 283, 212, 156, 117, 84 K) are shown.  The planetary surface gravity also changes among the models.  The three collections of black dots show quench pressures for log \kzz\ $=4$, 8, and 11. At depth, hotter profiles are clearly CO rich, while cooler profiles are CH$_4$-rich.
\label{ageprofs}}
\end{figure}

In the absence of tidal heating giant planet interiors inexorably cool as they age, meaning cooler interior adiabats and lower \tint\ values.  In the face of vertical mixing, we should expect atmospheric abundances to change then as well.  We examine the effect on a range of \emph{P--T} profiles for a Jupiter-like example (1 \mj, $3\times$ solar) at 0.15 AU in Figure \ref{ageprofs}.  The values of \tint\ are taken from every half-dex in planetary thermal evolution from an age of 10 Myr to 10 Gyr, yielding 7 models from \tint\ of 501 K to 84 K.  For moderately irradiated planets like these, the cooling of the interior has little effect on the upper atmosphere \citep{Sudar03}, but we should expect quite different atmospheric abundances when including vertical mixing.  The 3 sets of black dots in Figure \ref{ageprofs} show log \kzz\ of 4, 8, and 11.

\begin{figure}[htp]
\includegraphics[clip,width=1.0\columnwidth]{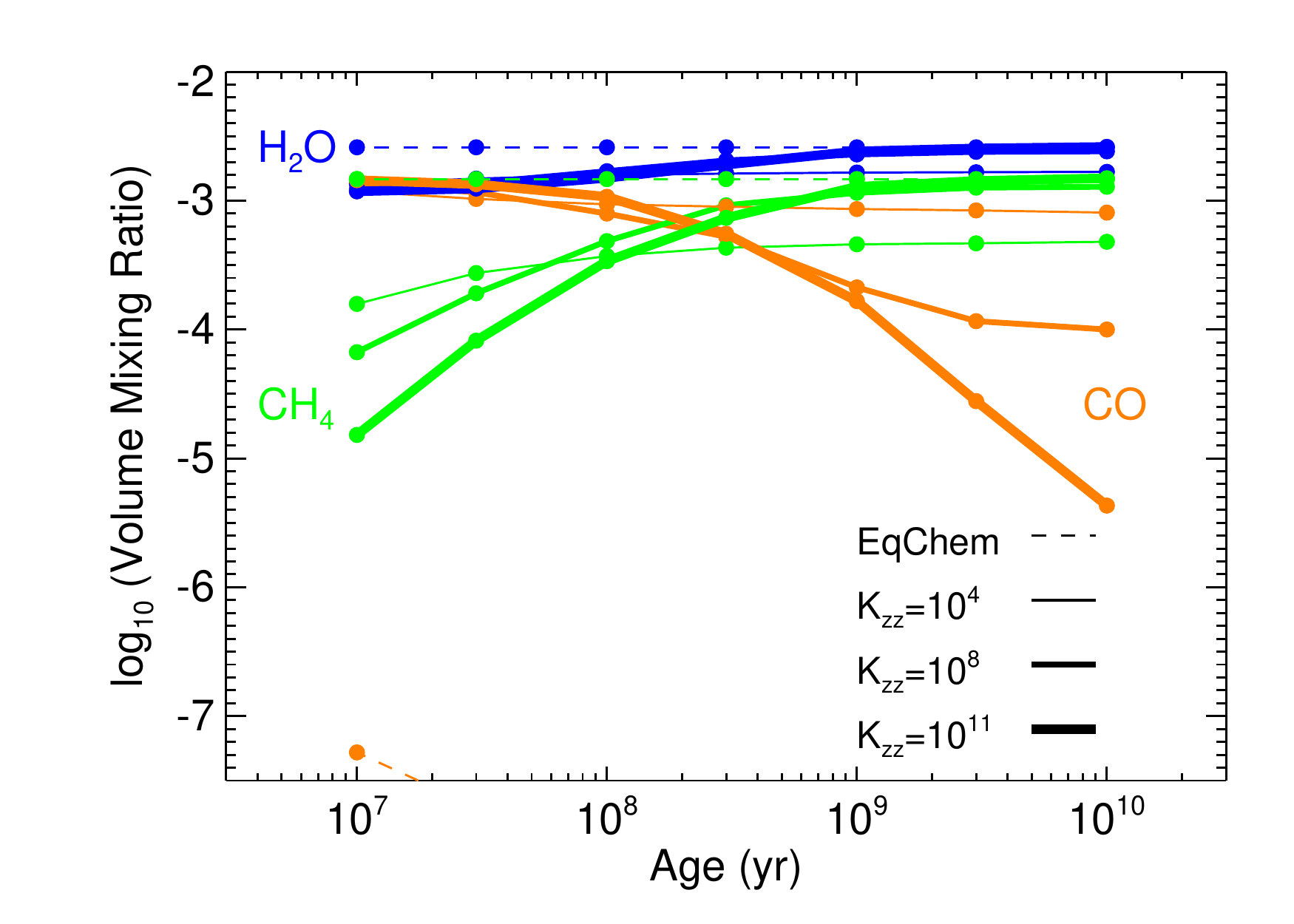}
\caption{Atmospheric abundances at 1 mbar as function of planetary age, for the \emph{P--T} profiles shown in Figure \ref{ageprofs}.  In equilibrium (dashed), the cooling of the planet's interior has \emph{no effect} on the atmospheric abundances, as the temperatures of the upper atmosphere are essentially constant, and the atmosphere could be CH$_4$-rich and quite CO-poor.  Modest vertical mixing (log \kzz =4) yields a much higher CO/CH$_4$ ratio, but abundances that again are essentially constant with time.  More vigorous mixing, from higher quench pressures, samples a much wider range range of CO and CH$_4$ abundances.  As the interior cools off the atmosphere transitions from CO-rich to CH$_4$ rich.
\label{agechem}}
\end{figure}

In Figure \ref{agechem} we examine the corresponding chemical abundances for equilibrium and the 3 values of vertical mixing strength, as a function of planetary age.  In equilibrium at 1 mbar, the atmosphere is CH$_4$ dominated, and the CO mixing ratio is nearly off the bottom of the plot.  However, even very modest vertical mixing (log \kzz\ $=4$, thin lines) changes the picture.  The atmosphere becomes modestly CO-dominated, and we lose essentially all sensitivity to the deeper atmosphere of the planet -- the abundances depend very little on \tint.  However with more vigorous vertical mixing, we see a picture emerge that has much in common with our understanding of non-equilibrium chemistry in brown dwarfs.  Higher \tint\ values and hotter interiors lead to more CO and less CH$_4$.  The plot shows a changeover from CO-dominated to CH$_4$-dominated at $\sim$~200 Myr, at a \tint\ value of $\sim 250$ K.  Again, this is generic behavior, as more massive objects would transition later in life (but at higher \tint\ values given their higher pressure photospheres and the positions of the CO and CH$_4$ iso-composition curves), and less massive objects earlier (but at higher \tint\ values, given their lower pressure photospheres).  While we expect building up a large sample of atmospheric spectra size a function of planetary age will be a challenge, it will be rewarding to have a statistical sample to compared to the typical several-Gyr-old systems.  This could yield important insights into planetary cooling history and the vigor of vertical mixing with age.

\subsection{N$_2$-NH$_3$ Transitions} \label{nitrogen}
Nitrogen chemistry is predominantly a balance between N$_2$ and NH$_3$, and has been explored and validated in the brown dwarf context \citep[e.g.,][]{Saumon00,Saumon03,Cushing06,Hubeny07,Zahnle14}.  N$_2$ is favored at high temperatures (and low pressures) while NH$_3$ is favored at low temperatures (and high pressures).  The transition from N$_2$ to NH$_3$ at cooler temperatures has a similar character to that of CO converting to CH$_4$, but it occurs at lower temperatures.  Understanding non-equilibrium nitrogen chemistry in brown dwarfs has typically been hampered by two constraints.  The first is that N$_2$, with no permanent dipole, has no infrared absorption features, unlike CO.  The second is that NH$_3$ iso-composition curves have slopes that lie nearly along interior H/He adiabats, meaning that one typically cannot assess a given atmosphere's quench pressure, as all pressures along the adiabat correspond to nearly the same NH$_3$ mixing ratio.

\begin{figure}[htp]
\includegraphics[clip,width=1.0\columnwidth]{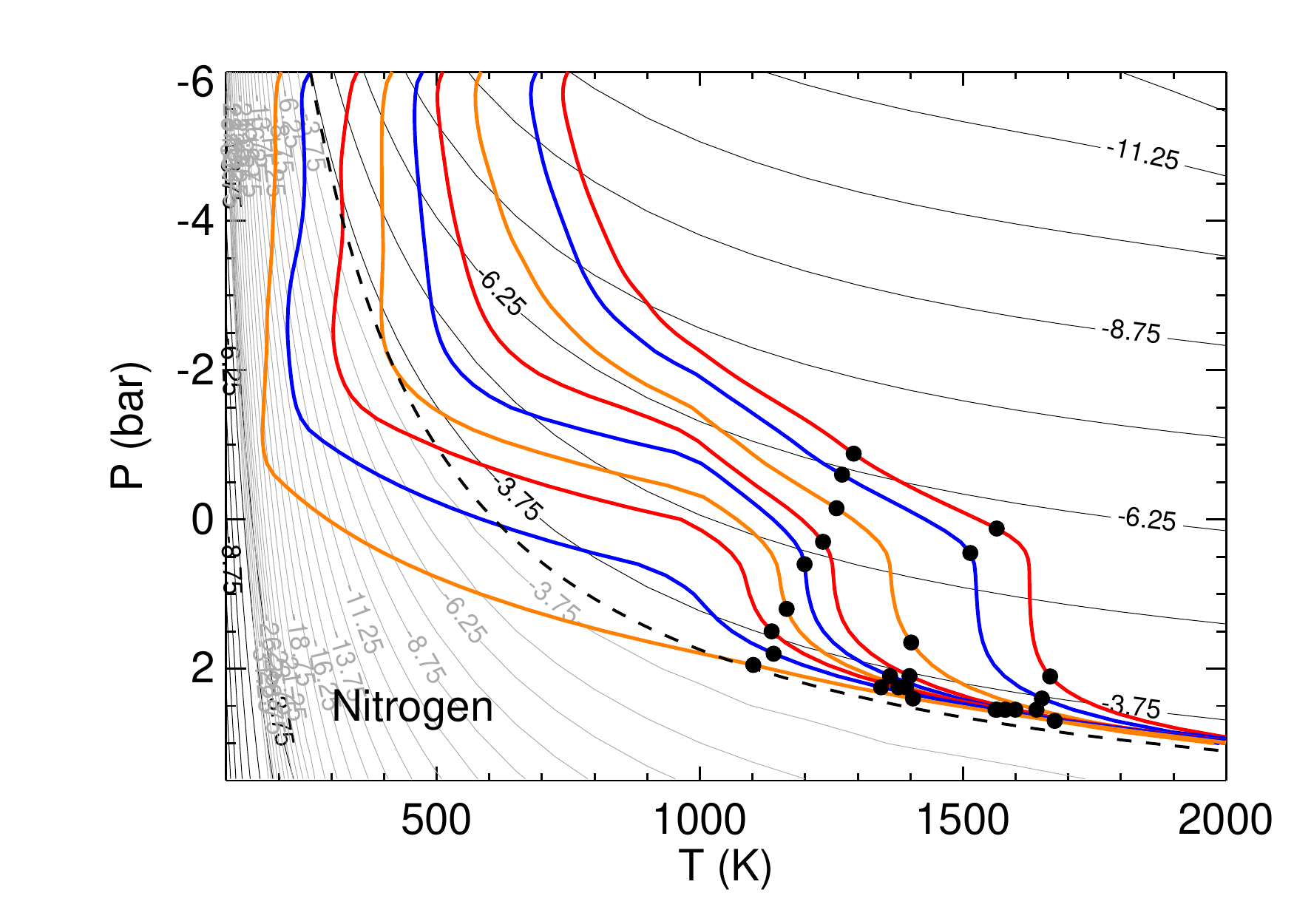}
\caption{Atmospheric \emph{P--T} profiles for old, Saturn-like planets (\tint$=75$K, $g=10$m s$^{-2}$, assuming $10\times$ metallicity.  The models are a 9 incident flux levels, at 0.06, 0.07, 0.1, 0.15, 0.2, 0.3, 0.5, 1, 2 AU from the Sun.  Three sets of black dots show the nitrogen quench pressure for log \kzz\ of 4, 8, and 11 cm$^2$ s$^{-1}$.  At higher pressures, note that the spread between all profiles is lessened, both in temperature, and in reference to the NH$_3$ (black) and N$_2$ (grey) abundance curves.
\label{nsaturnprofs}}
\end{figure}

However, in some sense irradiated planets have the advantage of having relatively more isothermal \emph{P--T} profiles, which can remain non-adiabatic to pressure of $\sim$~1 kbar.  And, \emph{if} these predominantly radiative atmospheres have \kzz\ values less than their mostly convective brown dwarf cousins, then it may be these more isothermal radiative parts of the atmosphere where one may quench the chemistry.  We can examine this with the same Saturn-like \emph{P--T} profiles we first examined in Figure \ref{saturnprofs}.  These profiles, but now with quench pressures for N$_2$-NH$_3$ chemistry \citep{Zahnle14}, are shown in Figure \ref{nsaturnprofs}.

Underplotted in black are curves of constant NH$_3$ abundance, falling off at higher temperature and lower pressure.  Underplotted in grey are curves of constant N$_2$ abundance, falling off at lower temperature and higher pressure.  A detailed look at Figure \ref{nsaturnprofs}, compared to Figure \ref{saturnprofs}, shows that the NH$_3$ iso-composition curves are more ``spread out'' than similar curves for CH$_4$, suggesting a more gradual change in nitrogen chemistry, with temperature, than for carbon.  As the chemical conversion times for N$_2 \rightarrow$ NH$_3$ are longer than for CO$\rightarrow$CH$_4$, the corresponding quench pressures for log \kzz$=4$, 8, and 11 cm$^2$ s$^{-1}$ are at somewhat higher pressures.  While for vigorous mixing (log \kzz$= 11$), all profiles converge to the same quench pressure (and hence changes in \teq\ across this range would yield no change in the NH$_3$ abundance, there are a broad ranges of N$_2$ and NH$_3$ mixing ratios for the log \kzz$=4$ and \kzz$=8$ cases.

\begin{figure}[htp]
\includegraphics[clip,width=1.0\columnwidth]{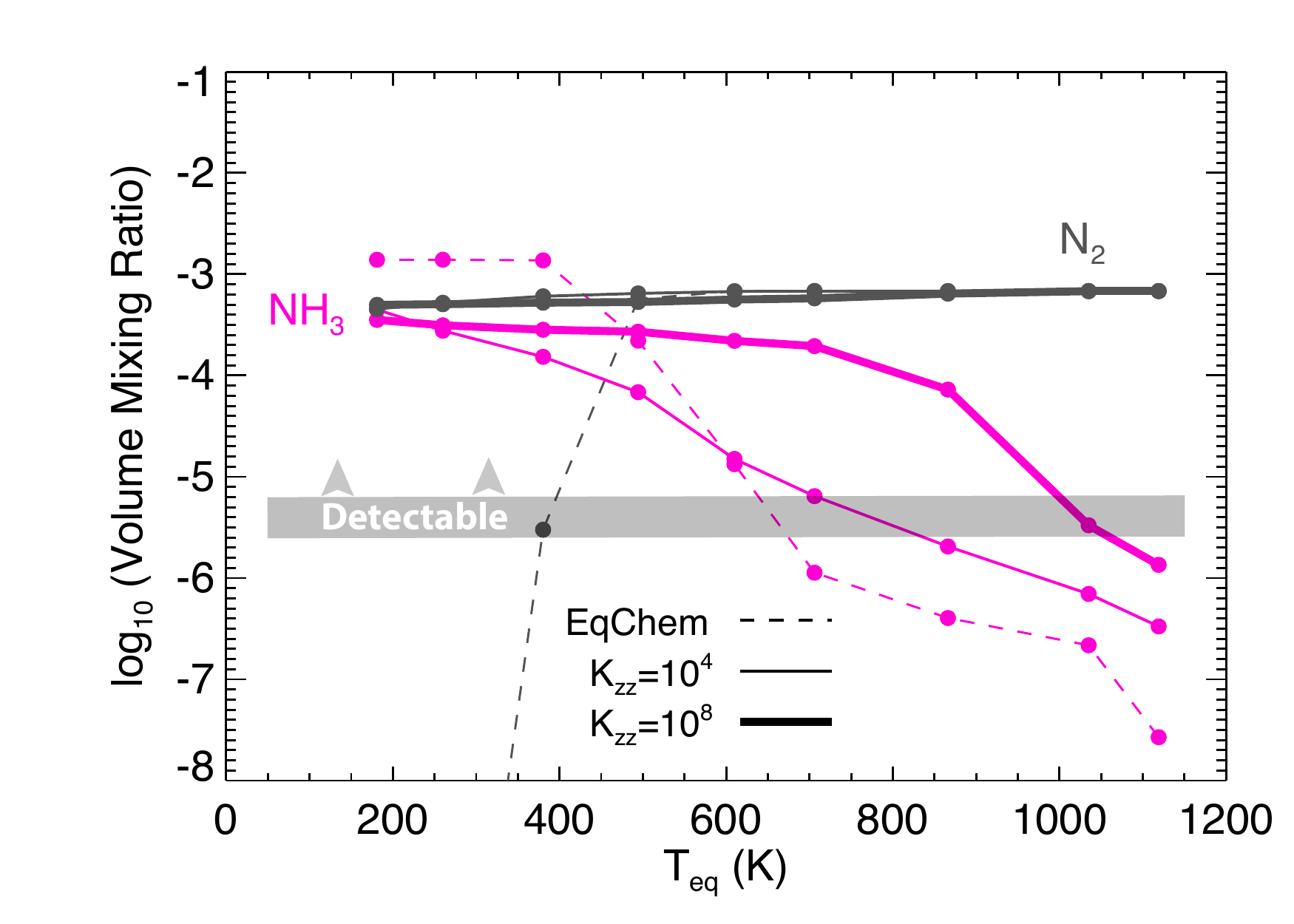}
\caption{The 9 \emph{P--T} profiles from Figure \ref{nsaturnprofs} are plotted at 9 \teq\ values across the x-axis, with chemical abundances along the y-axis.  ``EqChem" gives the nitrogen chemical equilibrium abundances at 1 mbar (dashed), while log \kzz\ $=4$ and 8 are shown as thin solid and thick solid, respectively.  In equilibrium, at \teq\ $\sim480$ K, the N$_2$ and NH$_3$ mixing ratios crossover, while for all models with vertical mixing, this crossover does not happen.  The more vigorous the vertical mixing, generally, the higher NH$_3$ mixing ratio, except for the coldest models.
\label{nsaturnchem}}
\end{figure}

Figure \ref{nsaturnchem} shows the mixing ratios of N$_2$ and NH$_3$ as a function of planetary \teq.  Equilibrium chemistry (at 1 mbar) shows a crossover from N$_2$-dominant to NH$_3$ dominant at around 475 K.  However, even sluggish vertical mixing keeps all of these atmospheres N$_2$ dominant, while also increasing the NH$_3$ mixing ratio for all \teq\ values $> 600$ K.  More vigorous mixing (log \kzz$=8$) further flattens the slope of the NH$_3$ curve, leading to relatively abundant NH$_3$ at essentially all \teq\ values, as expected from the grouping of most of the log \kzz$=8$ black dots in Figure \ref{nsaturnprofs}.  Across the entire phase space, the NH$_3$ mixing ratios are similar to those of CH$_4$ (see Figure \ref{saturnchem}), and are actually even \emph{higher} for NH$_3$ than for CH$_4$ for the higher \teq\ values.  This suggests that onset of detectable CH$_4$ is these planets should be accompanied by NH$_3$ as well -- one will not need to wait for particularly cold temperatures, compared to the brown dwarfs.  For those interested in determining the relative abundances of C, N, and O, to compare to Jupiter's values \citep{Wong04}, we note that in these models NH$_3$ never becomes the dominant nitrogen carrier compared to N$_2$, such that the nitrogen abundance determined from NH$_3$ would only be a lower limit.

\subsubsection{Effects of Planet Mass at a Given Age} \label{mass2}

\begin{figure}[htp]
\includegraphics[clip,width=1.0\columnwidth]{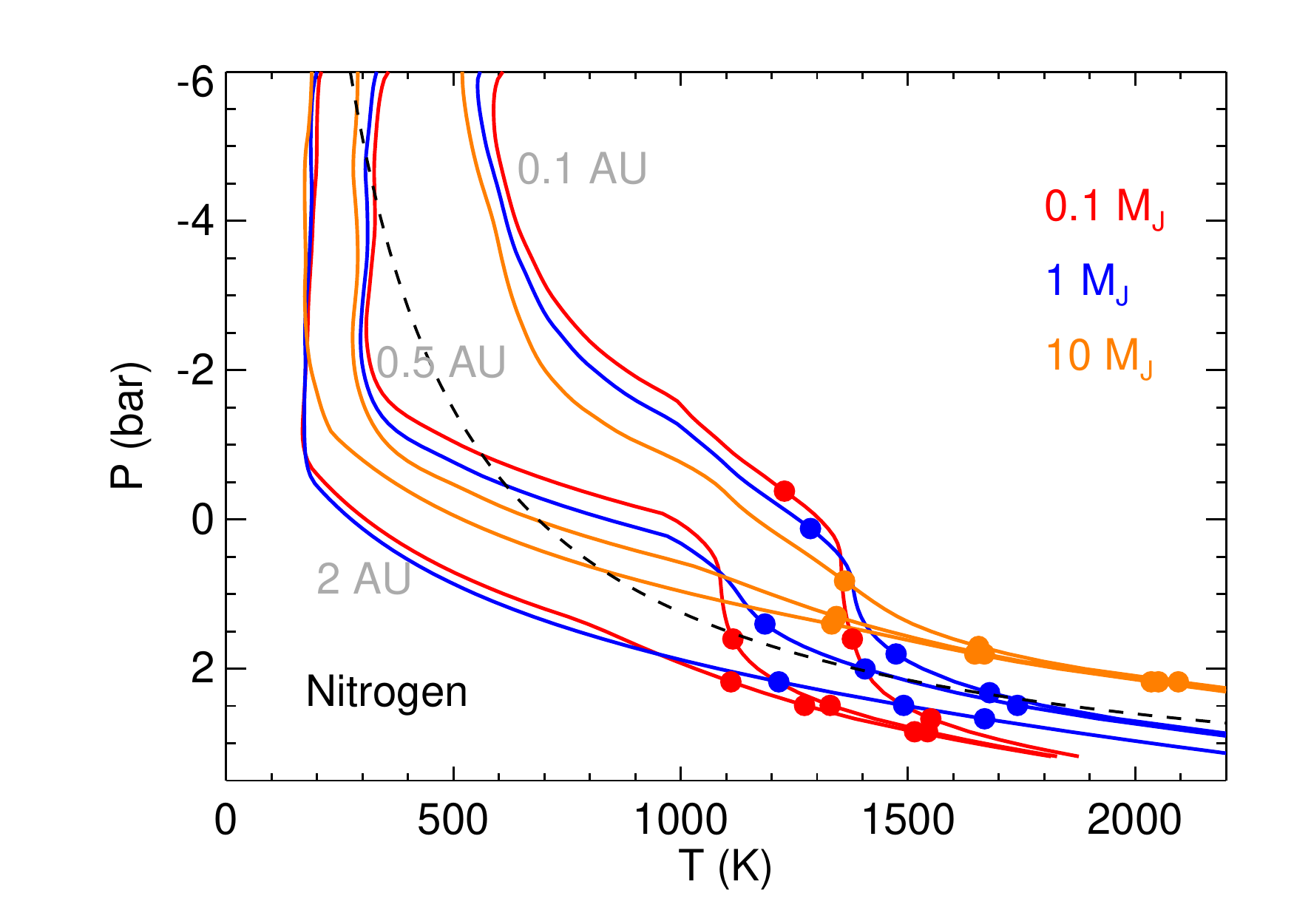}
\caption{Atmospheric \emph{P--T} profiles for 3-Gyr-old planets at 0.1 (red), 1 (blue), and 10 (orange) \mj, at $10\times$ solar.  The N$_2$/NH$_3$ equal-abundance curve is shown in black. The models are at 0.1, 0.5, and 2 AU from the Sun. The color-coded dots show the nitrogen quench pressure for log \kzz\ $=4$, 8, and 11.  Higher gravity models have higher pressure photospheres, but also have hotter interiors, which causes significant crossing of profiles.  The much larger scale heights for the low gravity models means greater physical distances for mixing, and hence, lower quench pressures.  Compared to Figure \ref{massprofs}, the nitrogen chemistry quench pressures are at higher pressures than for carbon chemistry.  For high gravity and/or cool models, the quench pressure is near or within the deep atmosphere adiabat.
\label{nmassprofs}}
\end{figure}

Previously, in Section \ref{mass1} and Figures \ref{massprofs} and \ref{masschem} we investigated the role that surface gravity and cooling history have for the planets.  Here, we examine the same profiles, but for nitrogen chemistry.  Figure \ref{nmassprofs} shows these sample \emph{P--T} profiles for the 0.1, 1.0, and 10 \mj\ planets, with log \kzz$=4$, 8, and 11.  Compared to the carbon example from Figure \ref{massprofs}, the quench pressures are higher.  For the high gravity (10 \mj) planet in particular, the quench pressure is within the deep atmosphere adiabat for log \kzz$=8$ and 11, and near it for log \kzz$=4$.  We might expect that the NH$_3$ abundance will change little with \kzz, similar to a brown dwarf case \citep{Zahnle14}.  The deeper one probes, the closer one comes to these adiabats, which lie nearly parallel to curves of constant NH$_3$ abundance.  Instead, the NH$_3$ mixing ratio is in some sense a probe of the current specific entropy of the adiabat, which could prove useful in constraining thermal evolution models.

\begin{figure}[htp]
\includegraphics[clip,width=1.0\columnwidth]{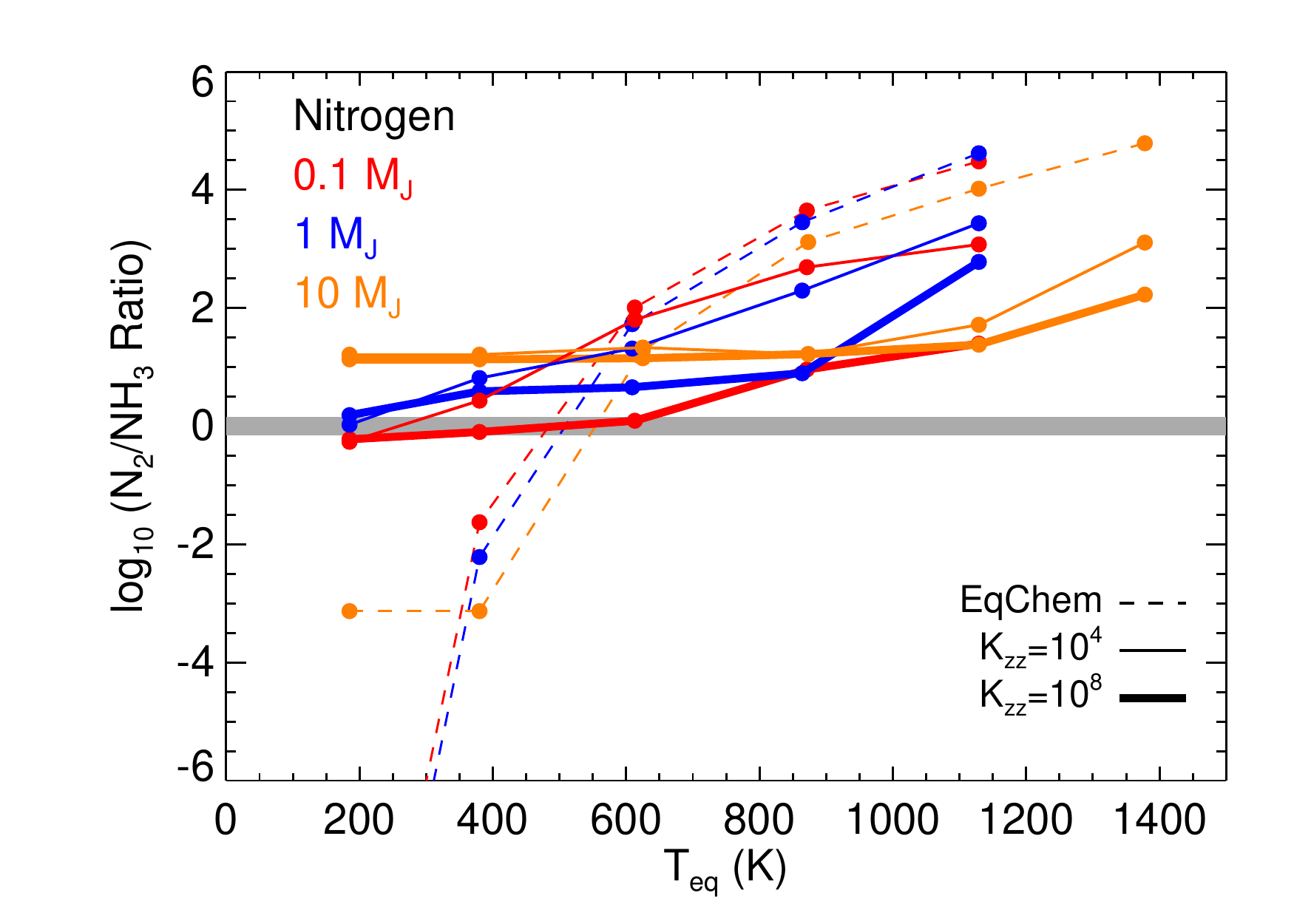}
\caption{The log of the N$_2$/NH$_3$ ratio for 5 values of \teq\ for 0.1, 1, and 10 \mj\ model planets, where a subset of the profiles are shown in Figure \ref{nmassprofs}.  In equilibrium (at 1 mbar), the transition \teq\ for CO/CH$_4$=1 (log=0, shaded grey) is at $\sim$ 420, 530, and 600 K, from low mass to high mass.  This is $\sim$400-500 K colder than the carbon chemistry transitions show in Figure \ref{masschem}.  However, vertical mixing essentially flattens the slopes of these curves, as one quenches from high pressure regions that lie on nearly the same adiabat, as shown in Figure \ref{nmassprofs}.  For all three model planets, NH$_3$ exists in detectable amounts for a wide swath of \teq\ values.
\label{nmasschem}}
\end{figure}

We can examine the N$_2$/NH$_3$ ratio as a function of \teq\ for these three planets in Figure \ref{nmasschem}.  The crossover \teq\ for nitrogen chemistry, in equilibrium, would be $\sim$550 K at 10 \mj, 500 K at 1 \mj, and 475 K at 0.1 \mj.  However, even modest vertical mixing dramatically changes this picture.  As the \teq\ decreases, the quench pressure falls near or into the deep atmosphere adiabat, even at low gravity.  On Figure \ref{nmassprofs} this manifests as the N$_2$/NH$_3$ ratio asymptoting to values that depend solely on the specific entropy of the adiabat, as one might have expected for the specific cases investigated for the Saturn-like planet in Figure \ref{nsaturnchem}.  Much like the brown dwarfs, at cool temperatures (and especially at high surface gravity) planets here are insensitive to \kzz.

\subsubsection{Effects of Planet Age at a Given Mass} \label{cooling2}
\begin{figure}[htp]
\includegraphics[clip,width=1.0\columnwidth]{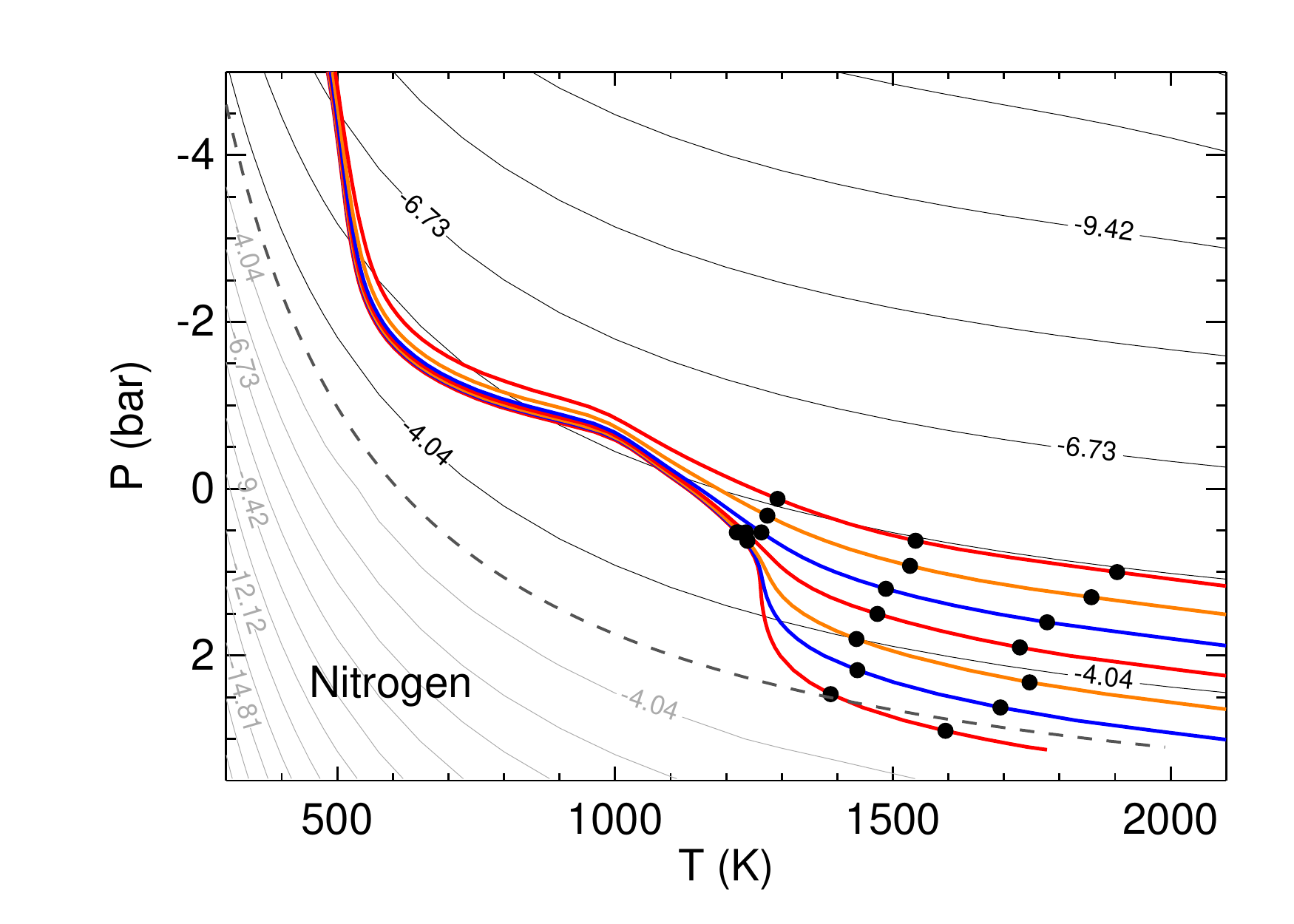}
\caption{Atmospheric \emph{P--T} profiles for a 1 \mj\ planet at 0.15 AU from the Sun, assuming $3\times$ solar metallicity.  Seven ages, every half dex from 10 Myr to 10 Gyr, with seven values of \tint\ (501, 383, 283, 212, 156, 117, 84 K, from Figure \ref{tracks}) are shown.  The three collections of black dots show nitrogen quench pressures for log \kzz\ $=4$, 8, and 11. At depth, all profiles are within the N$_2$ rich region of \emph{P--T} space, and the adiabats lie parallel to curves of constant NH$_3$ abundance.
\label{nageprofs}}
\end{figure}

Previously in Section \ref{cooling1} and Figures \ref{ageprofs} and \ref{agechem} we found that planet age, and hence, the cooling history and specific entropy of the interior adiabat, can have dramatic effects on the carbon chemistry.  Young planets would have quite different abundances (richer in CO) than older planets at the same \teq, all things being equal.  We can investigate the role of cooling history on the nitrogen chemistry with these same profiles.  In Figure \ref{nageprofs} we plot the 1 \mj\ profiles from 10 Myr to 10 Gyr, this time with the nitrogen quench pressures labeled. The figure is quite similar to \ref{ageprofs}, but with higher quench pressures, at hotter temperatures.  At log \kzz\ $=4$, the levels are in the radiative part of the atmosphere, but are relatively pinched together.  At log \kzz\ $=8$ and 11, we find all quench pressure in or very near the deep atmosphere adiabats.

\begin{figure}[htp]
\includegraphics[clip,width=1.0\columnwidth]{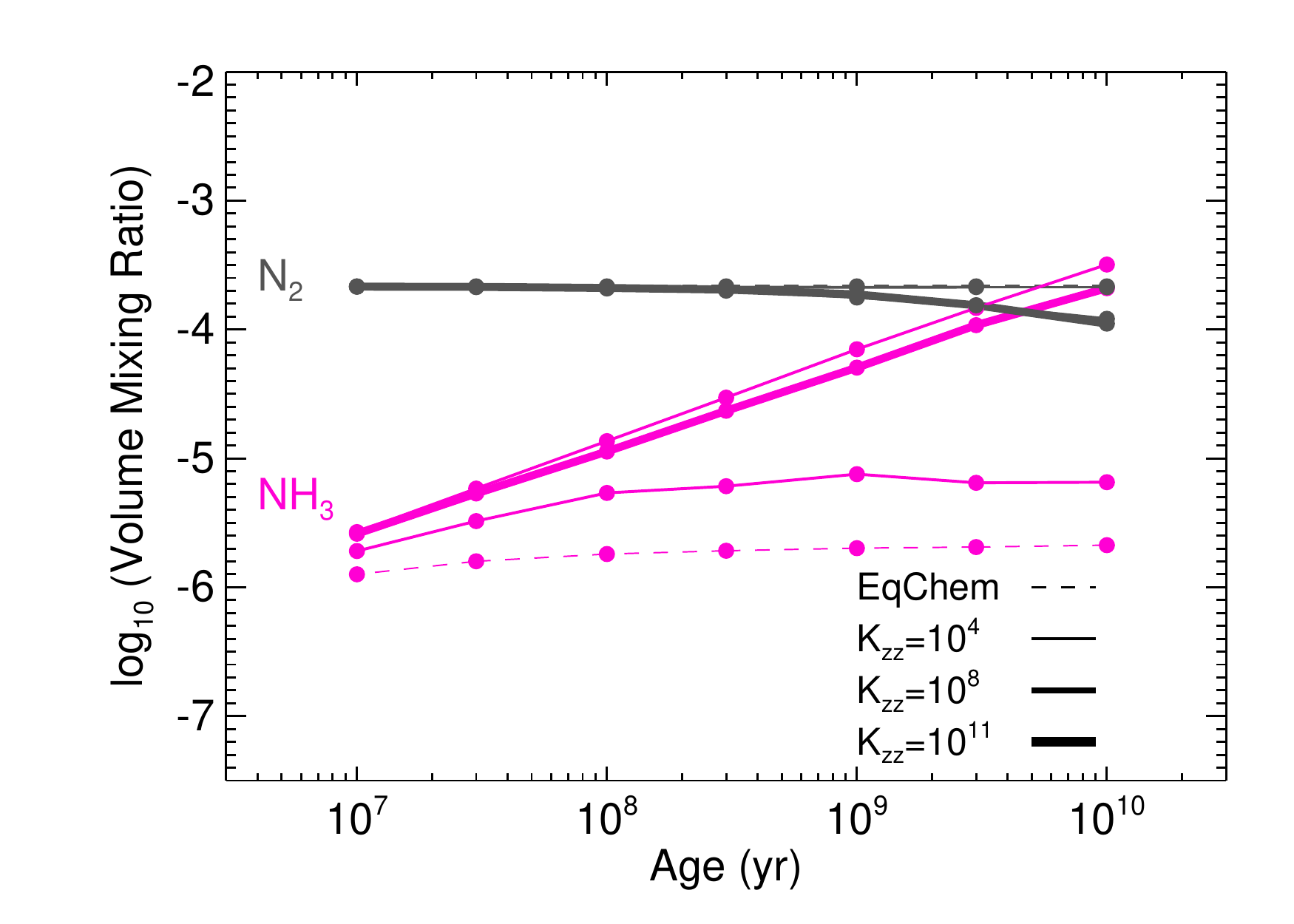}
\caption{Atmospheric N$_2$ and NH$_3$ abundances at 1 mbar as function of planetary age, for the \emph{P--T} profiles shown in Figure \ref{nageprofs}.  In equilibrium (dashed), the cooling of the planet's interior has almost no effect on the atmospheric abundances, as the temperatures of the upper atmosphere are essentially constant, and the atmosphere would be N$_2$ rich.  Modest vertical mixing (log \kzz\$=4) yields a slightly higher NH$_3$ abundance, but still essentially constant with time.  More vigorous mixing, from higher quench pressures (log \kzz\ or 8 and 11), samples progressively more NH$_3$-rich gas.  However, there is little sensitively in these models.
\label{nagechem}}
\end{figure}

The effect on the atmospheric mixing ratios of N$_2$ and NH$_3$, shown in Figure \ref{nagechem}, are quite straightforward, but different than that found for the carbon chemistry in Figure \ref{agechem}.  In equilibrium at 1 mbar, as the atmosphere changes negligibly in temperature, the NH$_3$ mixing ratio (dashed line) changes little with age.  The same is true at log \kzz\ $=4$, albeit it at a higher NH$_3$ abundance.  Since both the log \kzz\ $=8$ and 11 quench pressures sample the deep adiabat, which are nearly parallel NH$_3$ abundance curves, we find essentially the \emph{same} behavior of mixing ratio as a function of age, independent of (high) \kzz.  This is essentially the same as the well-understood brown dwarf behavior.

\subsection{Effect of a Mass-Metallicity Relation on Carbon and Nitrogen}
So far we have aimed, as much as possible, to investigate the physical and chemical effects of only altering one or two quantities at a time, including distance from the Sun, surface gravity, and \tint.  Atmospheric metallicity will also play an important role in altering these boundaries.  This chemistry has certainly be explored before, or a very wide range of compositions \citep[e.g.,][]{Moses13}.  In this section we attempt to explore a composition phase space, but in a more narrow sense.

It is strongly suggested from the bulk densities of transiting giant planets that there is a bulk ``mass-metallicity relation" for the planets \citep{Thorngren16}, with the lower mass giant planets being more metal-rich.  The effect of such a relation at \emph{atmospheric} abundances is not yet clear \citep{Kreidberg14b,Wakeford17,Welbanks19}, but there is such a relation in the solar system for carbon \citep[e.g.,][]{Atreya16}, and from standard models of core-accretion planet formation theory, albeit with a large spread \citep{Fortney13}.

\begin{figure}[htp]
\includegraphics[clip,width=1.0\columnwidth]{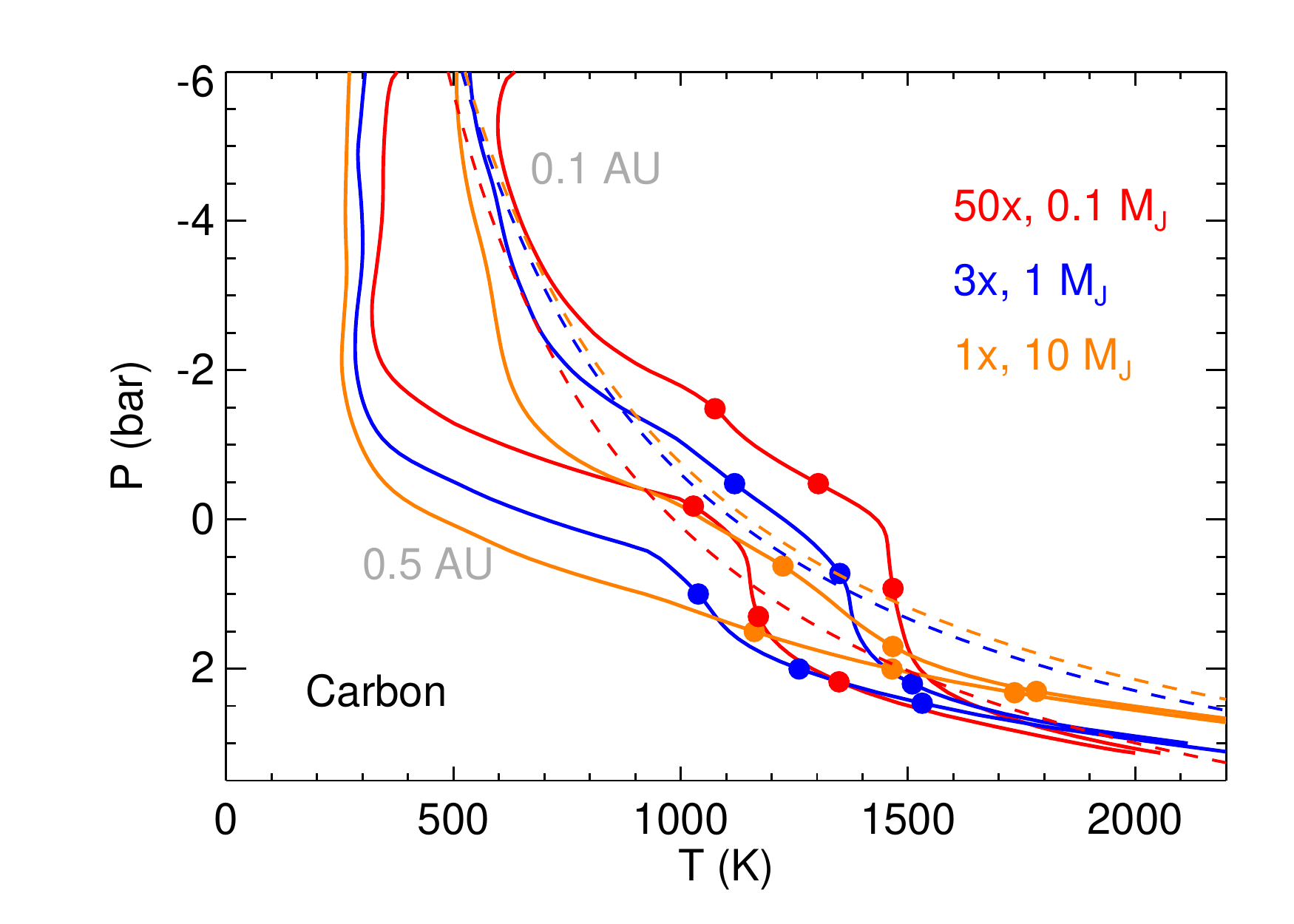}
\includegraphics[clip,width=1.0\columnwidth]{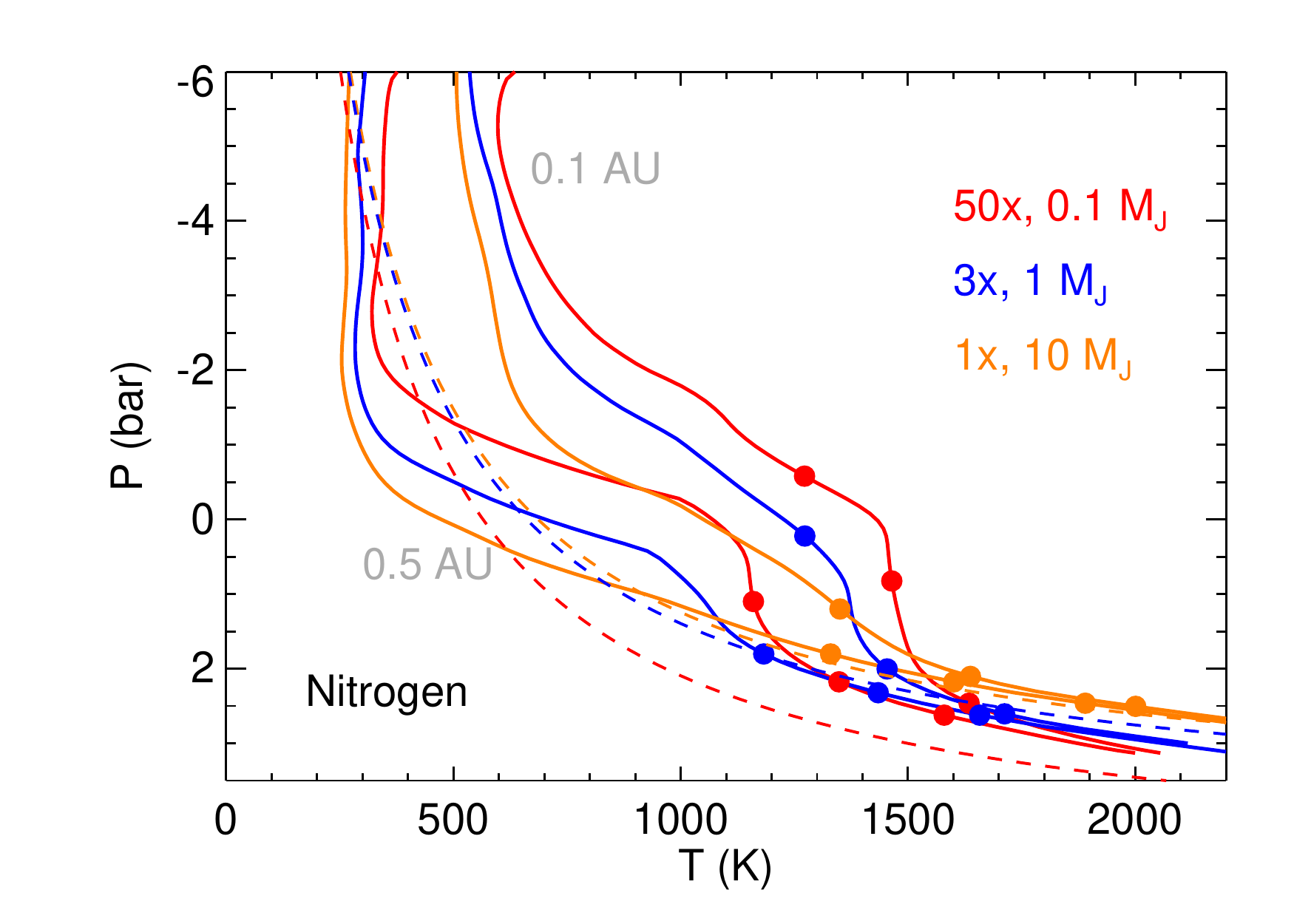}
\caption{Atmospheric \emph{P--T} profiles for 3-Gyr-old planets at 0.1 (red, $50\times$), 1 (blue, $3\times$), and 10 (orange, $1\times$) \mj.  The CO/CH$_4$ (upper) and N$_2$/NH$_3$ (lower) equal-abundance curves at these 3 metallicity values are shown in dashed curves with the same 3 colors.  The models are at 0.1 and 0.5 AU from the Sun. The color-coded dots show the quench pressures for log \kzz\ $=4$, 8, and 11 for carbon (upper panel) and nitrogen (lower panel).  The nitrogen chemistry quench pressures are at higher pressures than for carbon chemistry.  For high gravity and/or cool models, the quench pressure is near or within the deep atmosphere adiabat, in particular for nitrogen.
\label{metalpt}}
\end{figure}

For both the carbon and nitrogen chemistry discussed in Section \ref{mass1} and \ref{mass2}, for the 3 planet masses at 10$\times$ solar, we can examine how an increasing metallicity with lower planet masses may alter the previously examined trends.  Figure \ref{metalpt} shows \emph{P--T} profiles for planets at 0.5 and 2 AU from the Sun, with the upper panel showing carbon quench pressures and the lower panel nitrogen quench pressures.  The profiles themselves differ somewhat from those shown in Figure \ref{massprofs} and \ref{nmassprofs} as the models here use 50$\times$ solar (0.1 \mj), 3$\times$ solar (1 \mj), and 1$\times$ (10 \mj).  Since the plots use 3 different metallicities, we also show three different CO/CH$_4$ equal-abundance curves (dashed).

\begin{figure}[htp]
\includegraphics[clip,width=1.0\columnwidth]{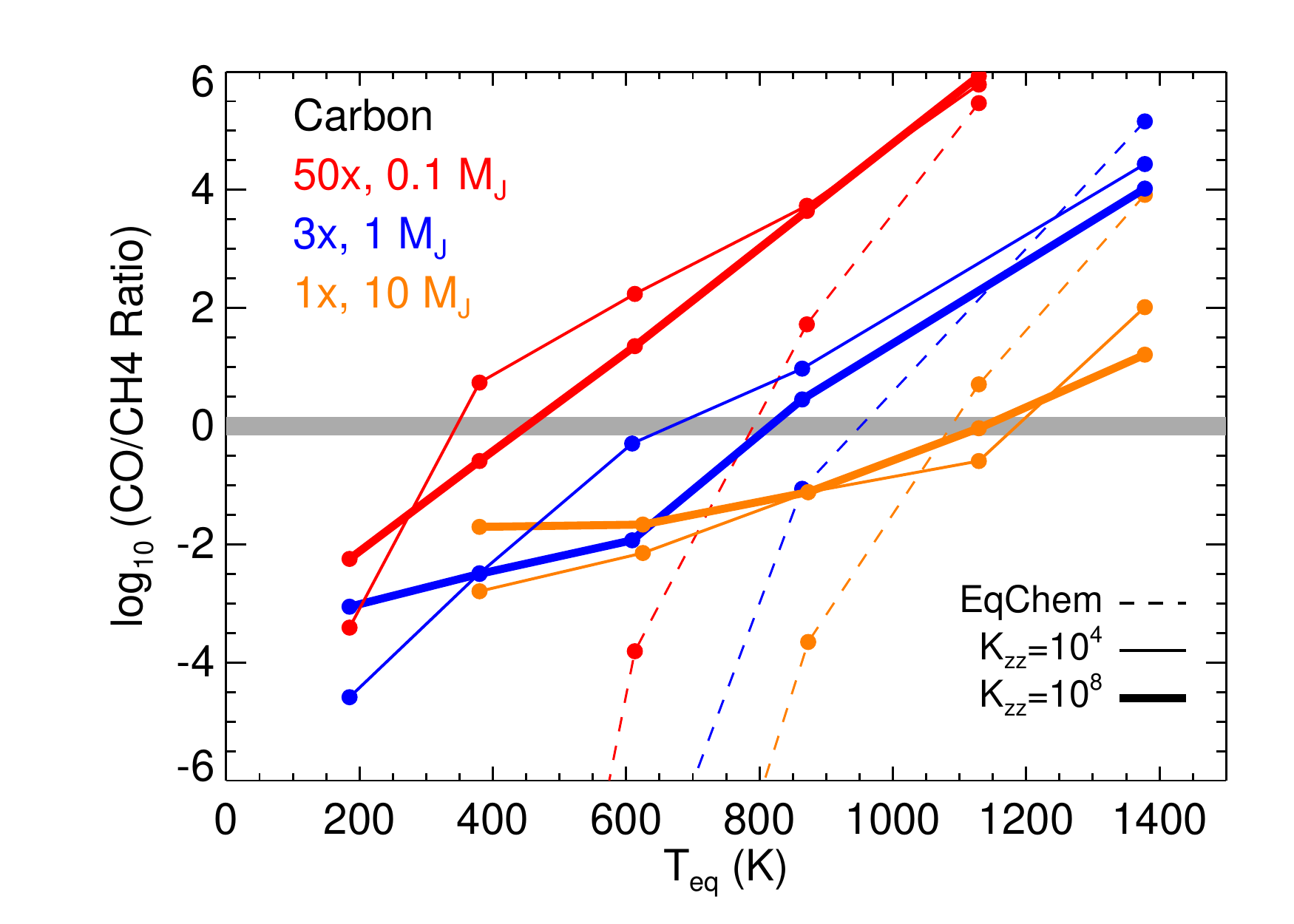}
\includegraphics[clip,width=1.0\columnwidth]{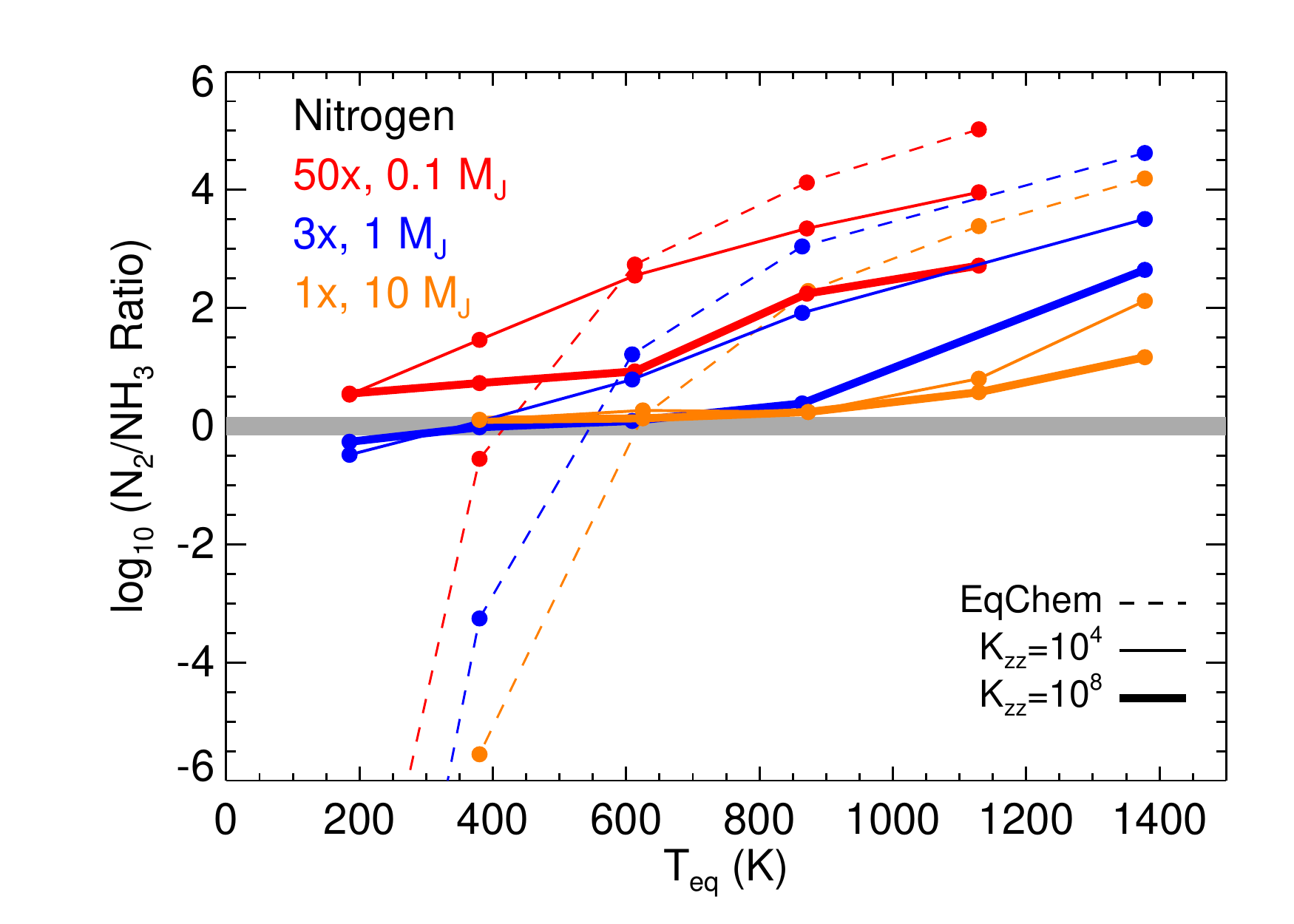}
\caption{The log of the CO/CH$_4$ ratio (upper panel) N$_2$/NH$_3$ ratio (lower panel) for 5 values of \teq\ for 0.1, 1, and 10 \mj model planets, where a subset of the profiles are shown in Figure \ref{metalpt}.  In equilibrium (at 1 mbar), the transition \teq\ for N$_2$/NH$_3$=1 (log=0, shaded grey, lower panel) is at $\sim$ 420, 530, and 600 K, from low mass to high mass.  This is $\sim$400-500 K colder than the carbon chemistry transitions in the upper panel.  For nitrogen in particular, vertical mixing essentially flattens the slopes of these curves, as one quenches from high pressure regions that lie on nearly the same adiabat, as shown in Figure \ref{metalpt}.  For all three model planets, NH$_3$ exists in detectable amounts for a wide swath of \teq\ values.
\label{metalchem}}
\end{figure}

Compared to our previous investigations into chemistry at 10$\times$ solar metallicity (Figures \ref{masschem} and \ref{nmasschem}), the two panels in Figure \ref{metalchem} show a much wider range of behavior.  At higher metallicity, the cooler models ``hang on'' to CO and N$_2$ to much cooler \teq\ values.  In equilibrium the carbon transitions would occur between 1100 and 700 K in these models. Even sluggish vertical mixing shows a large impact.  For instance, with more vigorous mixing (log \kzz$=8$), these three transition \teq\ values are $\sim$1100, 800, and 450 K.  %The \teq\ value for the onset of when CH$_4$ is ``detectable" is an important observational question, and one might suggest that CO/CH$_4$ ratio of $\sim$ 10$^4$ might be enough.  If so, then these \teq\ values would be $\sim$~1300 K for the 1 \mj\ planet, but only $\sim$~800 K for the 0.1 \mj\ planet.

We can examine the N$_2$/NH$_3$ ratio as a function of \teq\ for these three planets in Figure \ref{metalpt}.  The crossover \teq\ for nitrogen chemistry, in equilibrium, would be $\sim$600 K at 10 \mj, 530 K at 1 \mj, and 420 K at 0.1 \mj.  However, even modest vertical mixing dramatically changes this picture.  As the \teq\ decreases, the quench pressure falls near or into the deep atmosphere adiabat, even at low gravity.  On Figure \ref{nmassprofs} this manifests as the N$_2$/NH$_3$ ratio asymptoting to values that depend solely on the metallicity and the specific entropy of the adiabat, as one might have expected for the specific cases investigated for the Saturn-like planet in Figure \ref{nsaturnchem}.

\subsection{Putting it Together: The Onset of CH$_4$ and NH$_3$}
We can summarize, at least for the ``old" 3-Gyr planets that have been the baseline for many of calculations, the expected rise of detectable CH$_4$ and NH$_3$ abundances.  It is by now well-understood that for the atmospheres of brown dwarfs that the onset of CH$_4$ and NH$_3$ are well-separated in \teff-space.  Indeed, the rise of near-infrared CH$_4$ and NH$_3$ define the T and Y spectral classes, at $\sim$1300 K and $\sim$600 K respectively \citep{Kirkpatrick05,Stephens09,Line17}, although the much stronger mid-IR bands can appear at 1700 K (CH$_4$ at 3.3 $\mu$m) and 1200 K (NH$_3$ at 10.5 $\mu$m).

\begin{figure}[htp]
\includegraphics[clip,width=1.0\columnwidth]{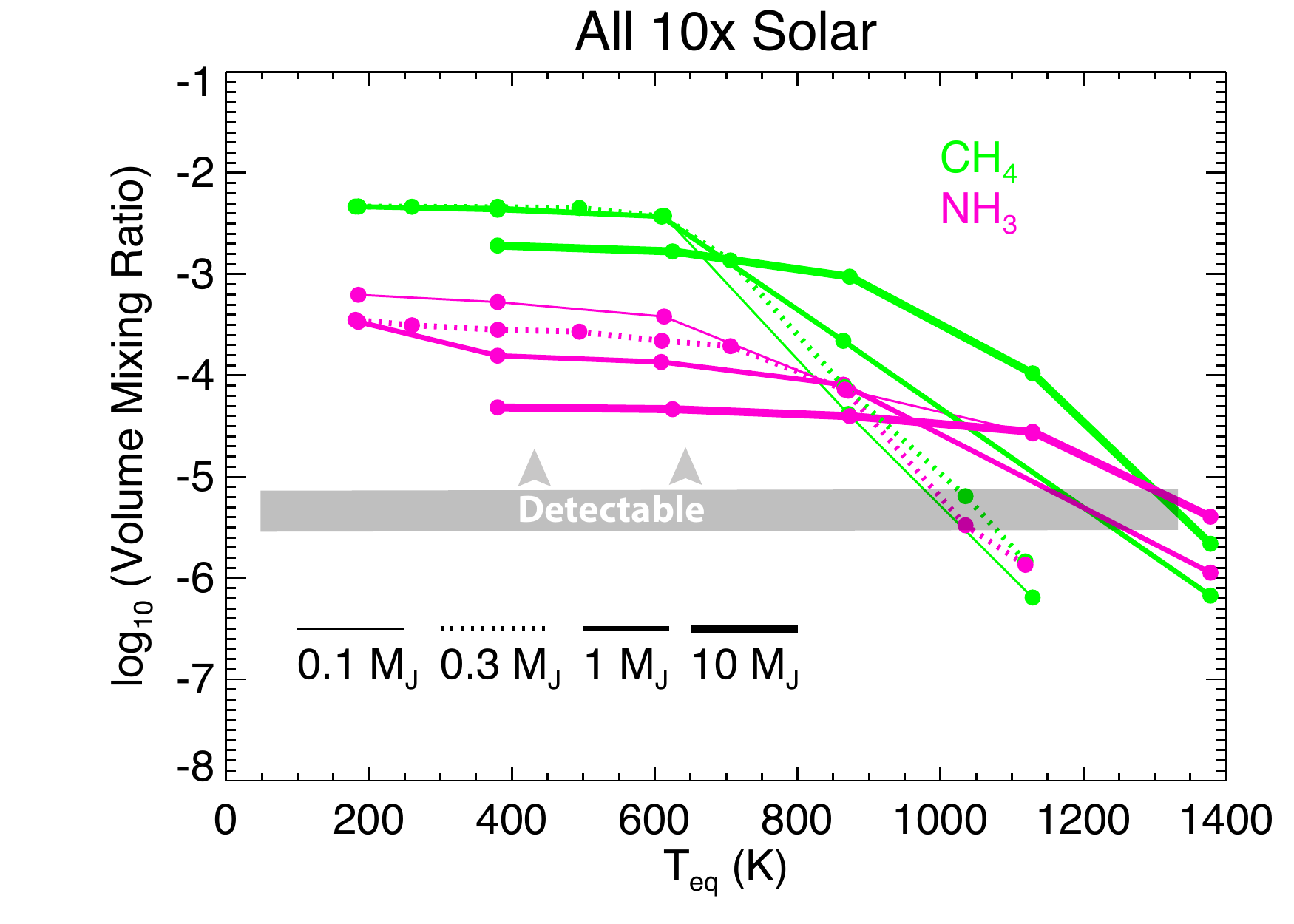}
\includegraphics[clip,width=1.0\columnwidth]{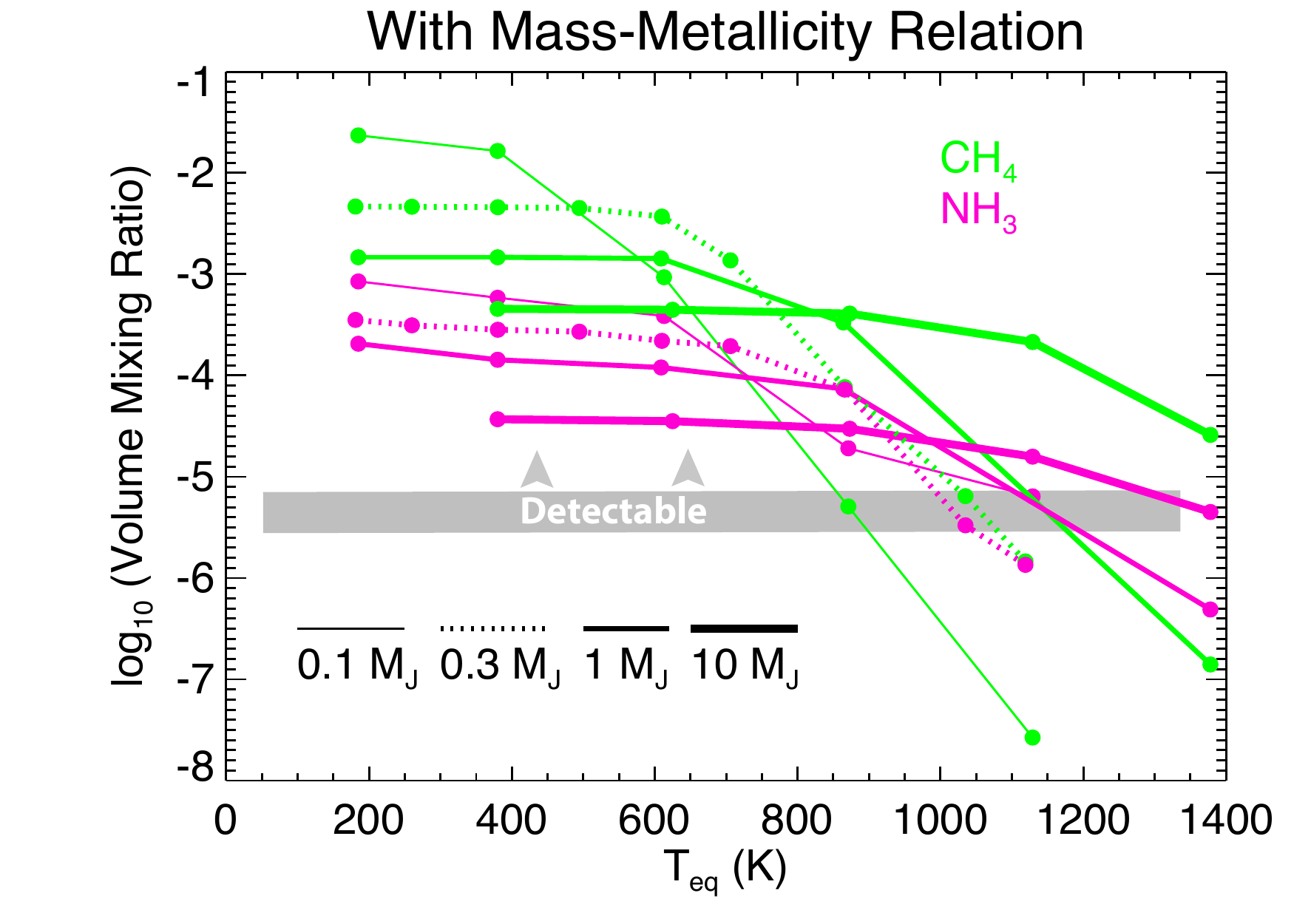}
\caption{The log of the CH$_4$ and NH$_3$ mixing ratios as a function of \teq\ for models at  0.1, 0.3, 1, and 10 \mj\ model planets at an age of 3 Gyr. The upper panel shows calculations where 10$\times$ solar abundances are used for all models, while the lower panel assumes the mass-metallicity relation (50, 10, 3, and 1$\times$ solar) for the 4 masses, respectively.  For the range of models, and unlike in brown dwarfs, the onset of NH$_3$ is nearly coincident with the onset of CH$_4$, and for the lower masses ($<0.3$ \mj), NH$_3$ onset occurs for warmer \teq\ values than CH$_4$.  In this figure log \kzz\ $=8$ is assumed.
\label{onset}}
\end{figure}

However, significantly different \emph{P--T} profiles of irradiated giant planets leads to much different behavior.  This is shown in Figure \ref{onset}, both for planets at a fixed 10$\times$ solar metallicity (top panel) and for planets that use the notional mass-metallicity relation (bottom panel), with both panels using log \kzz\ of 8.  For the higher gravity planets with a large thermal reservoir in their interior, the giant planet behavior is at least similar to that of brown dwarfs, with CH$_4$ coming on for \teq\ a few hundred K hotter for the 1$\times$ solar case at 10 \mj\ (bottom panel).  However, beyond that example, a different and richer behavior, driven mostly by the altered temperature structure of irradiated planets, is seen.  For all other example planets in both panels, CH$_4$ and NH$_3$ onset is at a similar \teq, and at the higher metallicities (bottom panel) NH$_3$ can arise at \emph{warmer} \teq\ values than CH$_4$.

Figure \ref{onset} is in some ways the central prediction of the paper, albeit for a relatively constrained example, as we describe at some length in the Discussion section.  The oddly shaped and radiative \emph{P--T} profiles lead to an expectation of significantly different behavior than that already known for brown dwarfs.   

\subsection{Cloud Formation and Cold Traps}
A lesson well-learned from observations of transiting planet atmospheres to date is that clouds and hazes can readily obscure molecular absorption features. This has typically been thought of as a hindrance. However, early work in this field suggested that the atmospheres of giant planets could potentially be classified based on the presence or absence of clouds \citep{Marley99,Sudar00,Sudar03}. In the end, it seems likely that some mixture will be true -- in some ways clouds will help us understand temperature structures and transport in these atmospheres, but will also obscure features due to atoms and molecules.

However, it seems clear that the role of clouds will not be a simple function of \teq, as cloud condensation curves can be crossed at a variety of pressures.  At a low pressure, perhaps little condensible material will exist.  At a high pressure, perhaps all cloud material in an optically thick cloud will be below the visible atmosphere. These effects will depend on the shape of the atmospheric \emph{P--T} profile, and hence on the specific entropy of the adiabat (which depends on planet mass and age), in addition to the role of atmospheric metallicity (more metals means more cloud-forming material), and even the spectral type of the parent star, which can also alter profile shapes, as discussed below.

\begin{figure}[htp]
\includegraphics[clip,width=1.0\columnwidth]{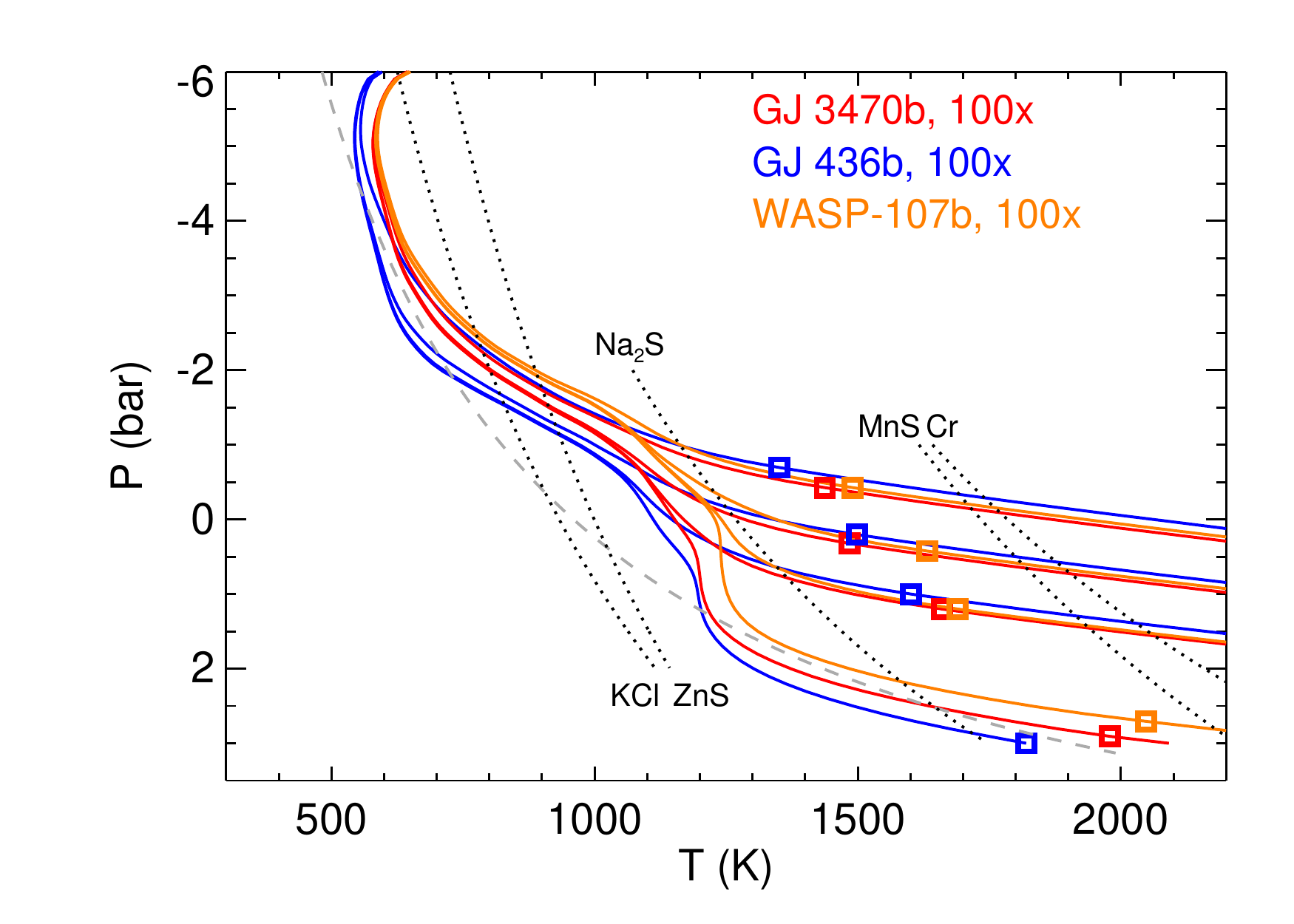}
\caption{Atmospheric \emph{P--T} profiles for planets GJ 436b, GJ 3470b, and WASP-107b all at $100 \times$ solar abundances, taken from Figure \ref{3profiles}.  Black dashed curves are for cloud condensation for various elements from \citet{Morley12}.  For each planet, 4 interior adiabats are shown, for the case of no tidal heating (coolest), and $Q=10^6$, $10^5$ and $10^4$, from cooler to warmer.  Colored squares show the radiative-convective boundary depth.  Tidal heating can push cloud formation of Na$_2$S, MnS, and Cr, out of the deep atmosphere, into the visible atmosphere.
\label{cloudstrio}}
\end{figure}

In some ways this topic is beyond the scope of the paper, which is focused on 1D models, but we can motivate that there will be a diversity in behavior at a given planetary \teq\ with plots that focus on \emph{P--T} profiles and condensation curves.  First we will examine our trio of warm Neptunes, GJ 436b, GJ 3470b, and WASP-107b.  In Figure \ref{cloudstrio} we replot the same \emph{P--T} profiles from Figure \ref{3profiles}, with chemical information removed, but now including radiative-convective boundary depths (RCBs) with squares, and condensation curves for potential cloud-forming materials.  These ``cooler" clouds, for planets cooler than the hot Jupiters, have been studied in \citet{Morley12, Morley13}.  Note, however, that \citet{Gao20} have suggested that most of these cloud species (save KCl) may not nucleate and form.  \citet{Lee18} suggest that Cr, KCl, and NaCl (instead of Na$_2$S) will form across this temperature range.  These predictions can be corroborated by future detailed spectroscopic observations of brown dwarfs and planets.

The KCl and ZnS cloud bases move little with or without tidal heating, as the upper atmospheres change little.  The Na$_2$S cloud base, however, can move dramatically.  Without tidal heating, the cloud base would be around $\sim$300 bars in all three planets.  However, for tidal heating with $Q=10^4$, the Na$_2$S cloud base moves to $\sim$0.1 bar, in the visible atmosphere.  A similar effect is seen for MnS and Cr.

\begin{figure}[htp]
\includegraphics[clip,width=1.0\columnwidth]{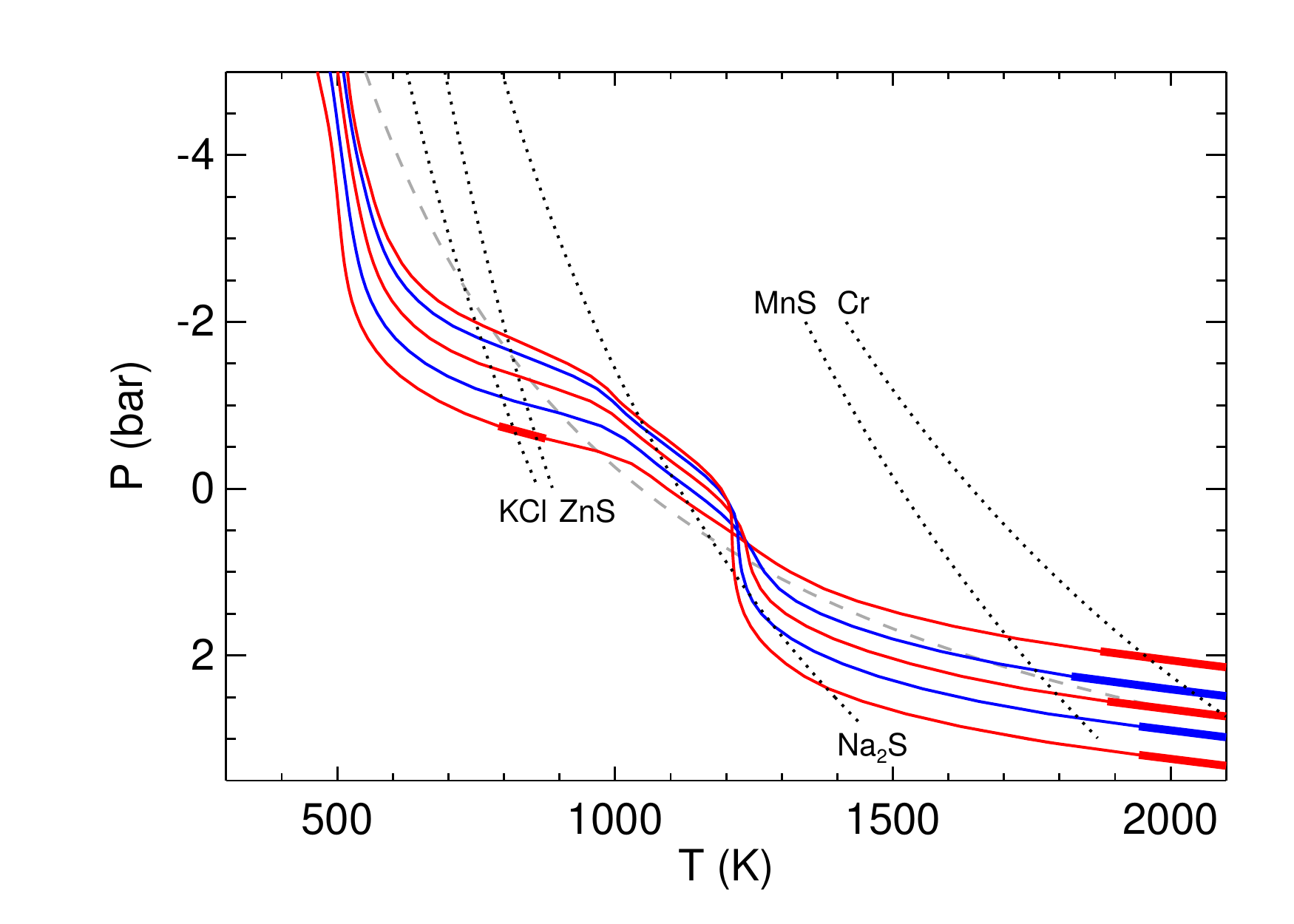}
\caption{Model pressure-temperature profile for a $10\times$ solar atmosphere at 0.15 AU from the Sun, The five profiles from Figure \ref{gravity} show (alternating red and blue) five values of \tint, at 52, 77, 117, 182, and 333 K, with respective surface gravities $g$=5.8, 9.8, 24, 65, and 225 m s$^{-2}$.  Thicker parts of the profiles show convective regions.  Note that the specific entropy of the deep atmosphere adiabat can move the location of the Na$_2$S cloud into the visible atmosphere (base 1 bar for the highest gravity model) or a depth (base at 300 bar in the lowest gravity model).  The high gravity model also has a detached convective zone (coincidentally) at the location of ZnS and KCl condensate formation.
\label{cloudtint}}
\end{figure}

We have previously investigated generic Saturn-like-planet \emph{P--T} profiles at 0.15 AU from the Sun.  Figure \ref{cloudtint} shows the same profiles that were explored in Figure \ref{gravity}, now with a focus on RCBs and cloud condensation, rather than chemical abundances.  The interface between these profiles and condensation depends strongly on surface gravity.  For instance, the denser, higher pressure photosphere of the highest gravity models yields a detached convective zone near 0.2 bar, coincidentally at the region of ZnS and KCl clouds, which is not seen in the lower gravity models.  Potentially more vigorous mixing here could lead to thicker clouds and larger particle sizes.  If these profiles were calculated at greater orbital distances, yielding cooler atmospheres, all would develop this detached convective zone \citep{Fortney07a}.  The Na$_2$S case is also interesting for these profiles.  The cloud base is found in the deep atmosphere for the two higher gravity models, but at a few tenths of bar in the three lower gravity models.  This clearly shows that at a given \teq, the depth of cloud formation can be significantly impacted by temperature of the deep atmosphere, which is mitigated by the interior cooling.  One could readily imagine other examples where the cloud formation depth is affected by planetary age, at a given mass, as is seen in brown dwarfs and self-luminous imaged planets.

\section{Discussion}  \label{discussion}
\begin{figure}[htp]
\includegraphics[clip,width=1.0\columnwidth]{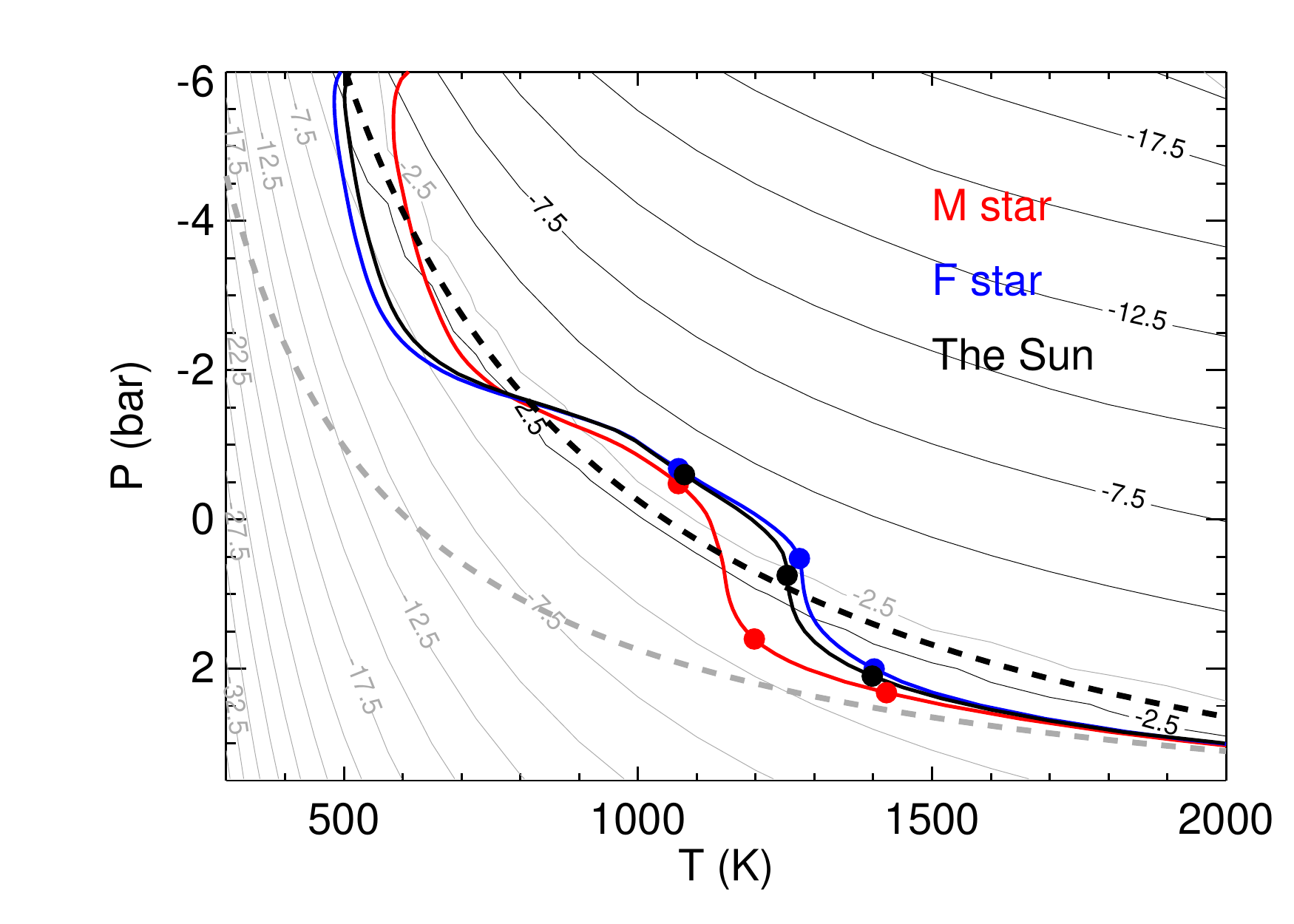}
\caption{Atmospheric \emph{P--T} profiles for three planets with the same incident stellar flux.  For the profile in black, the planet is at 0.15 AU.  In red is a profile with the GJ 436b parent star (type M2.5), while in blue it is the WASP-17b parent star (type F4).  Here log \kzz values of 4, 8, and 11 are shown as upper, middle, and lower set of color dots, respectively.  Large temperature differences are particularly seen at at 1-100 bars, the relevant quench pressures for log \kzz=8, which straddles the CO/CH$_4$ equal abundance curve (dashed black).  The N$_2$/NH$_3$ equal abundance curve is shown in dashed gray, for reference.
\label{stars}}
\end{figure}

We wish to stress that the calculations shown here are only a starting point, and we have considered only what we believe will be the 1st order effects.  In the interest of brevity we have not considered several additional factors that could or will play important roles in further altering predicted temperature structures and atmospheric abundances.  We describe these here:
\begin{enumerate}
\item We have elected not to self-consistently recalculate the atmospheric \emph{P--T} profiles for each value of \kzz.  The altered atmospheric abundances in turn alter the radiative-convective equilibrium profile, as has been explored by several authors, with and without stellar irradiation \citep{Hubeny07,Drummond18,Phillips20}.  In particular \citet{Drummond18}, for HD 189733b and HD 209458b, found differences in the \emph{P--T} profile of up to 100 K.  For the arguments presented here, tripling or quadrupling the number of plotted \emph{P--T} profiles (one for every \kzz) would distract from the main point, particularly given the large uncertainly today in the \kzz\ profiles.  Additionally, including the cloud species discussed here would alter \emph{P--T} profiles and chemical transitions \citep{Mola20}.

\item We have assumed a constant value of \kzz\ with height.  Mixing length theory is an important guide to \kzz\ in convective regions, but it is not yet clear how \kzz\ transitions at the radiative-convective boundary, in particular given the 3D nature of atmospheric mixing.  Three-dimensional GCM runs may be a guide for particular planets of interest.  Work to date has suggested that as one moves deeper, to higher pressures in the radiative regions, that \kzz\ should decrease.  This may lead to a ``quench bottle neck" of less vigorous mixing just above the RCB.

\item Our models are 1D, however 3D effects have been shown to be important in understanding atmospheric abundances.  As has previously been demonstrated \citep{Cooper06,Agundez14,Drummond18b,Drummond20}, non-equilibrium chemistry is affected by day-night temperature differences in addition to vertical mixing.  Day-night effects \emph{may} be minimized for these relatively cooler planets, compared to the hot Jupiters, as day-night temperature differences are expected to be more modest at cooler temperatures \citep{Lewis10,PerezBecker13}.

\item Non-solar ratios of elemental abundance ratios are likely to occur.  As has been extensively modeled over the past decade, planet formation processes can drive atmospheres towards higher or lower C/O ratios, depending on the formation location and the relative accretion of solids and gas \citep[e.g.,][]{Oberg11,Madhu14,Mordasini16,Espinoza17}.  More recently, the role of the nitrogen N$_2$ ice line as a site of planet formation \citep{Piso16,Bosman19,Oberg19} and altered N/O and N/C ratios in giant planet atmospheres \citep{Cridland20} has been investigated.  Previous radiative-convective atmospheric calculations have shown that an altered C/O ratio can alter \emph{P--T} phase space of major chemical transitions \citep[e.g.,][]{Madhu11,Molliere15}.

\item Photochemistry will further alter atmospheric abundances.  The nonequilibrium abundances that we find, based on timescale arguments, are merely the ``raw materials'' for further chemical reactions \citep{Zahnle09,Zahnle09b,Moses11,Moses13,Venot20}.  It is well known that CH$_4$ in the solar system can be readily photolyzed, and the destruction of CH$_4$ may make it less easily observed, while increasing the abundances of other hydrocarbons, along with photochemical hazes.  We note that signs of hazes may already be seen in the transmission spectra of the cool transiting giant planet population \citep{Gao20}.

\item  A range of parent star spectral types will be relevant across the planetary population.  Moving from hot stars to cool stars, the peak of the stellar spectral energy distribution moves to redder wavelengths, and the temperature of the incoming radiation field is more similar to that of the planetary atmosphere, leading to more isothermal temperature structure \citep{Molliere15}, as shown in Figure \ref{stars}.  The range from hotter to cooler parent stars certainly spans at least the range from F to M.  Temperature differences of $\sim$150 K are seen at at 1-100 bars, the relevant quench pressures for log \kzz=8, which straddles the CO/CH$_4$ equal abundance curve.  Interestingly, this could be a very nice probe of \kzz, as for this example, as much lower and much higher \kzz values, the profiles converge back to similar CO/CH$_4$ abundances.

\item A range of planetary eccentricities can impact the timescale arguments made here, as well as drive tidal heating.  The thermal response of the planetary atmospheric temperatures, and hence chemistry, depends on the planetary orbit.  The timescale over which the atmosphere heats up and cools off due to the eccentric orbit will compete with the timescales \tmix\ and \tchem\ that we have explored here.  This idea was previously explored for  highly eccentric hot Jupiters by \citet{Visscher12}, but a new study that focuses on cooler planets appears to be warranted.  Tidal heating from the interior, as shown for planets GJ 436b, GJ 3470b, and WASP-107b in Section \ref{tides}, should be a relatively common process, particularly for the ``in-between" planets that are not so close that they will have circularized quickly, and are not so far tides do not affect the energy budget.  Tidal heating should then be investigated for any particular target of interest.  Assessing the eccentricity of a given planet may be difficult, if radial velocity data is sparse, or if a secondary eclipse is not detected.

\item The radius-inflation mechanism that affects hot Jupiters may still operate in the cooler planets we investigate here.  Since \ct{Thorngren18} and \ct{Thorngren19}, found no strong evidence for the mechanism affecting planets cooler than \teq$<1000$ K, we have used standard thermal evolution models that lack additional heating.  However, modest additional internal heating could warm the deep atmosphere, with only small effects on the observed radius vs. incident flux distribution, which would be currently undetectable in the planetary population.  And any ``residual" radius inflation power could be important for the Saturn- and Neptune-class planets, whose interiors would be expect to cool of significantly in the absence of additional power.  This would lead to lower CH$_4$/CO and NH$_3$/N$_2$ ratios at a given \teq, compared to our calculations, and could be an important probe of temperatures in the deeper atmosphere.

\end{enumerate}

\section{Conclusions}
Through a straightforward implementation of 1D radiative-convective model atmospheres and non-equilibrium chemistry, we have shown that atmospheric abundances of C-, N-, and O-bearing molecules in warm transiting planets will show a diverse and complex behavior.  This behavior will depend strongly on the cooling history of the planet, such that a planet's mass, age, parent star spectral type, and any ongoing tidal dissipation can lead to atmospheric abundances that differ from planet to planet at the same level of incident stellar flux. 

Non-equilibrium chemical abundances may then serve as a tool to probe the deeper atmosphere, similar to work recently begun for very cool brown dwarfs \citep{Miles20}.  For the three Neptune-class planets discussed in Section \ref{tides} (GJ 436b, GJ 3480b, and WASP-107b), we suggest that ongoing eccentricity damping tidally heats the deep atmospheres of the planets.  This raises temperatures by several thousand degrees and drives strong convective mixing, which dramatically decreases the CH$_4$/CO ratio in the visible atmosphere.  This may play the dominant role in understanding their observations to date.

The more isothermal shape of \emph{P--T} profiles in irradiated planets, compared to brown dwarfs, leads to the expectation that planetary behavior will differ strongly compared to brown dwarfs.  Perhaps most strikingly, the onset of detectable CH$_4$ and then NH$_3$ should occur at very similar \teq\ values, and for the Saturn-masses and below, a reversal compared to brown dwarf behavior, where NH$_3$ is seen at warmer temperatures than CH$_4$.  We have also shown that N$_2$ will dominate over NH$_3$ over a wide range of temperatures and ages, such than bulk nitrogen abundances determined from NH$_3$ will only be lower limits.

To discover the underlying physical and chemical trends for these atmospheres, it would likely be the most straightforward to look for \emph{trends at a given mass and age}.  For instance, in mature planetary systems (say, Gyr+), the Jupiter-mass planets around Sunlike stars at \teq$<1000$ K would all be expected (barring tidal heating) to have \tint\ values of $\sim$100 K.  One could expect to see a trend of increasing CH$_4$ abundance with lower \teq, with CH$_4$ becoming dominant at 800 K, as in Figure \ref{masschem}.  Note, however, that this potential trend could readily be disguised by mixing planets with a range of masses into one's sample, as shown in that same figure.  We reiterate that it is not yet known how diverse the atmospheric metallicities of those planets may be, and how that may change with planetary mass, which would also add scatter to any trend.

While retrievals to constrain atmospheric abundances and temperature structures \citep[see][for a review]{Madhu18} are likely up to the task for determining abundances in planetary transmission and emission, these findings can only properly be interpreted \emph{within the context of the physical characteristics of the planet and its environment}.  In particular, since we find that \tint\ can play a significant role in altering abundances, retrievals that utilize deep atmospheric temperatures that are guided by thermal (and/or tidal) evolution models, and aim to retrieve the quench pressure depth in addition to molecular mixing ratios, may yield the most robust results.  The role of planetary structure modeling, thermal evolution modeling, and physics-driven 1D and 3D models, to complement retrieval, are be essential to interpreting observations.

\acknowledgements
JJF wishes to dedicate this paper to Adam P. Showman, a great scientist, long-time collaborator, and good friend.  The authors thank Caroline Morley and Jacob Bean for insightful discussions during the course of this project.  JJF, MSM, MRL, RSF, and RL acknowledge the support of NASA Exoplanets Research Program grant 80NSSC19K0446.

%\bibliography{sample63}{}
%\bibliographystyle{aasjournal}

%\bibliography{output}{}
%\bibliographystyle{aasjournal}

%\bibliographystyle{aasjournal}
%\bibliography{references.bib}

\end{document}